\documentclass[sn-mathphys]{sn-jnl}

\jyear{2022}%

\theoremstyle{thmstyleone}%
\theoremstyle{thmstyletwo}%

\theoremstyle{thmstylethree}%

\begin{document}

\title[Barrow holographic dark energy with Granda-Oliveros cut-off]{Barrow holographic dark energy with Granda-Oliveros cut-off}


\author{\fnm{A.} \sur{Oliveros}}\email{alexanderoliveros@mail.uniatlantico.edu.co, \hyperlink{https://orcid.org/0000-0002-6796-1784}{ORCID}}

\author{\fnm{M. A.} \sur{Sabogal}}\email{msabogal@est.uniatlantico.edu.co,\hyperlink{https://orcid.org/0000-0002-9109-9442}{ORCID}}

\author*{\fnm{Mario A.} \sur{Acero$^{*}$}}\email{marioacero@mail.uniatlantico.edu.co,
\hyperlink{https://orcid.org/0000-0002-0835-0641}{ORCID}}

\affil{\orgdiv{Programa de F\'isica}, \orgname{Universidad del Atl\'antico}, \orgaddress{\street{\qquad\qquad\qquad Carrera 30 No.~8-49}, \city{Puerto Colombia}, \state{Atl\'antico}, \country{Colombia}}}


\abstract{A study on the effects of implementing the Granda--Oliveros infrared cutoff in the recently introduced Barrow Holographic Dark Energy model is presented, and its cosmological evolution is investigated. We find how the deformation parameter, $\Delta$, affects the evolution of $H(z)$, and that from this model it is possible to obtain an accelerated expansion regime of the universe at late times. We also observe that increasing $\Delta$ causes a transition of the EoS parameter from quintessence to phantom regimes. In addition, we show that the model can be used to describe the know eras of dominance. Finally, after studying the stability of the proposed model, a fit of the corresponding parameters is preformed, utilizing the measurements of the expansion rate of the universe, $H(z)$. The best fit of the parameters is found to be $(\alpha,\, \beta,\, \Delta) = (1.00^{+0.02}_{-0.02},\,0.69^{+0.03}_{-0.02},\,0.000^{+0.004}_{-0.000})$ at $1\sigma$ C.L., for which the Bekenstein-Hawking relation is favored.}

\keywords{Holographic, Barrow, Dark Energy, Granda--Oliveros cutoff}



\maketitle

\section{Introduction}\label{sec1}
Nowadays, there is a general consensus among cosmologists about of the existence the so-called dark energy (DE), which is behind the accelerated expansion of the universe at late times. This phenomenon has been supported by a huge amount of cosmological observations, but it was originally reported in 1998 by two independent research groups, the High-z Supernova Search Team and the Supernova Cosmology Project \cite{SupernovaSearchTeam:1998fmf, SupernovaCosmologyProject:1998vns}. Yet, there is currently no adequate solution to the DE problem and its explanation from fundamental theories of physics is still unknown.

Within the wide spectrum of proposals to handle the DE problem, there is a hypothesis known as the \emph{Holographic principle} \cite{tHooft:1993dmi, Susskind:1994vu}, which plays an important role in quantum gravity. The main idea behind the holographic principle is that the entropy of a system does not scale with its volume, but with its surface area \cite{Bousso:1999xy}. Inspired by this proposal, Cohen et al.~\cite{Cohen:1998zx}, suggested that, in a quantum field theory, a short distance cutoff is related to a long distance cutoff due to the limit set by the formation of a black hole, namely, if $\rho$ is the quantum zero-point energy density caused by a short distance cutoff, the total energy in a region of size $L$ should not exceed the mass of a black hole of the same size, hence, $L^3\rho\leq LM_{pl}^2$. The largest (infrared cutoff) $L_{_{IR}}$ allowed is the one saturating this inequality, thus 
\begin{equation}\label{eq:DEdensity}
\rho=3M_{p}^2c^2L_{_{IR}}^{-2},
\end{equation}
where $c$ is an arbitrary parameter, and $M_{p}$ is the reduced Planck mass. This holographic consideration has been widely applied in cosmology, especially for the description of late time dark energy era, and it is commonly known as holographic dark energy (HDE) (see \cite{Wang:2016och} for an extensive review). From this point of view, the infrared cutoff, $L_{_{IR}}$, has a cosmological origin and the authors of \cite{Nojiri:2005pu, Nojiri:2017opc,Nojiri:2021iko,Nojiri:2020wmh} introduced the most general form for this cutoff (so-called generalized HDE), which includes some combination of the FRW parameters, e.g., the Hubble constant, the particle and future horizons, the cosmological constant and the universe life-time --if finite.

A new proposal that has recently caught the attention of the community is the so-called Barrow holographic dark energy (BHDE) \cite{Saridakis:2020zol}, which has its roots in the idea introduced by Barrow in \cite{Barrow:2020tzx}, inspired by the COVID-19 virus illustrations. Barrow proposed that due to quantum gravitational effects, the black hole Bekenstein-Hawking entropy \cite{Bekenstein:1973ur, Hawking:1975vcx} must be modified introducing a fractal structure for the horizon geometry:
\begin{equation}\label{eq:BHentropy}
S_B=\left(\frac{A}{A_0}\right)^{1+\frac{\Delta}{2}},
\end{equation}
where $A$ is the standard horizon area, $A_0$ the Planck area, and $\Delta$ is the deformation parameter, which indicates the amount of the quantum gravitational deformation effects to the horizon structure. Notice that, for $\Delta=0$, the standard Bekenstein-Hawking entropy is recovered, while $\Delta = 1$ corresponds to the most intricate fractal structure. When the BHDE is implemented using the holographic principle and the new Barrow entropy proposal, Eq.~(\ref{eq:BHentropy}), the resulting holographic energy density becomes 
\begin{equation}\label{eq:BarrowDE_rho}
\rho_{\Lambda}=CL^{\Delta-2},
\end{equation}
where $C$ is a parameter having dimensions $[L]^{-2-\Delta}$ \cite{Barrow:2020tzx}. Clearly, for $\Delta=0$, Eq.~(\ref{eq:BarrowDE_rho}) reduces to the standard HDE, i.e., $\rho_{\Lambda}=3M_{p}^{2}c^2L^{-2}$, where $C=3M_{p}^{2}c^2$ with $c^2$ the corresponding model parameter.

The infrared cutoff $L$ used in \cite{Barrow:2020tzx} was the future event horizon $R_h$ and from this, the author shows that under this scenario it is possible to describe the thermal history of the universe, with the sequence of matter and dark-energy eras. Additionally, the new Barrow exponent, $\Delta$, significantly affects the dark-energy equation of state (EoS), and according to its value it can lead the EoS to lie in the quintessence regime, in the phantom regime, or to experience the phantom-divide crossing during its evolution. Furthermore, in \cite{Anagnostopoulos:2020ctz} the authors
used observational data from Supernovae (SNIa) Pantheon sample, as well as from direct measurements of the Hubble parameter from the cosmic chronometers (CC) sample, to impose constraints on the Barrow holographic dark energy  scenario. It is important to notice that the BHDE has been widely studied in the literature, e.g., \cite{Mamon:2020spa, Dabrowski:2020atl, Srivastava:2020cyk, Pradhan:2021cbj, Adhikary:2021xym, Leon:2021wyx, Asghari:2021bqa, Jusufi:2021fek, P:2021edm, Nojiri:2021jxf, Nojiri:2022aof, Huang:2021zgj, Luciano:2022pzg, Wang:2022hun, DiGennaro:2022ykp, Nojiri:2022dkr, Farsi:2022dvd}.

It is well known that a model using $R_h$ as the length scale predicts the observed current acceleration of the universe; however, this kind of models also display the causality problem, i.e., DE at present appears to depend on the future evolution of the scale factor, which in turns violates causality \cite{Li:2004rb}. Hence, here we analyze the cosmological evolution at late times in the framework of the BHDE model using the Granda-Oliveros (G-O) cutoff which, in addition to the square of the Hubble parameter, also includes the time derivative of the Hubble parameter, namely,
\begin{equation}\label{eq:GOcutoff}
L_{IR}=(\alpha H^2+\beta\dot{H})^{-1/2},
\end{equation}
where $\alpha$ and $\beta$ are arbitrary dimensionless parameters. This cutoff was proposed by the authors of Refs.~\cite{Granda:2008dk,Granda:2008tm}, considering  purely dimensional arguments, and succeed to avoid the causality problem. Also, we perform an analysis aimed to fit the parameters in the BHDE model with the G-O cutoff, using updated measurements from the dynamics of the expansion of the universe, $H(z)$. Recently in \cite{P:2021edm}, the authors have considered the Barrow holographic dark energy with Granda-Oliveros length as IR cutoff, but in a scenario where the BHDE is a dynamical vacuum, and taking into account an interaction between the matter and dark energy sectors.

The article is organized as follows: 
In Section \ref{sec_Model} we describe the details of the proposed model and the resulting evolution equations for the most important quantities.
The change of these quantities with the redshift is presented also shown in this Section, looking at the effects of different and interesting values of the parameters; the stability of the model under consideration is also studied in this section.
The constraints on the parameter imposed by the requirement of stability allow us to set specific ranges which are then implemented in Section \ref{sec_Fitting} to perform a fit of the parameters considering measurements of the Hubble parameter as a function of the redshift, $H(z)$. 
Finally, the conclusions are posed in Section \ref{sec_Conclusions}.

\section{The model}\label{sec_Model}
As stated in Sec.~\ref{sec1}, we are implementing the BHDE density in Eq.~(\ref{eq:BarrowDE_rho}) with the G-O IR cutoff, Eq.~(\ref{eq:GOcutoff}). In such a model, the holographic DE density becomes
\begin{equation}\label{eq:ModelDE_rho}
\rho_{\Lambda}=3 M_{p}^{2} \left(\alpha H^{2}+\beta \dot{H}\right)^{1-\frac{1}{2} \Delta}. 
\end{equation}
In this case, contrary to the original HDE model with the G-O cutoff (for $\Delta=0$),  $\alpha$ and $\beta$ are parameters with dimension $[L]^{\frac{2\Delta}{\Delta-2}}$, for the sake of preserving the dimensions of $\rho_{\Lambda}$. Furthermore, note that in this modified version of $\rho_{\Lambda}$ the parameter $C$ in Eq.~(\ref{eq:BarrowDE_rho}) has been replaced by $3M_{p}^{2}$ (considering $c=1$), since in our case, the role played by $c$ in Eq.~(\ref{eq:BarrowDE_rho}) via $C=3 M_{p}^{2}c^2$ is performed now by  $\alpha$ and $\beta$.

Since observations suggest that the universe is homogeneous and isotropic at large scales, we consider a flat Friedmann-Robertson-Walker (FRW) geometry with metric
\begin{equation}
d s^{2}= - dt^{2} + a^{2}(t)\delta_{i j} d x^{i} d x^{j},
\end{equation}
where $a(t)$ is the scale factor, and the content of the universe at large-scale is taken to be a perfect fluid. The first Friedmann equation for a universe made of non relativistic matter, radiation and DE is given by
\begin{equation}\label{eq:Model_FriedmannEq1}
3 M_{p}^{2}H^{2}= \rho_m+\rho_r+\rho_\Lambda.
\end{equation}
Considering these components as barotropic fluids ($p_i=w_i\rho_i$, with $i=m, r$), matter ($w_m = 0$) and radiation ($w_r = 1/3$) behave as 
\begin{equation}\label{eq:Model_densities}
\rho_m=\rho_{m_{0}}a^{-3},\qquad \rho_r=\rho_{r_{0}}a^{-4},
\end{equation}
which together with Eq.~(\ref{eq:ModelDE_rho}) for $\rho_{\Lambda}$, Eq.~(\ref{eq:Model_FriedmannEq1}) takes the form
\begin{equation}\label{eq:Model_FriedmannEq}
H^{2}= \Omega_{m_{0}}H_{0}^{2} a^{-3} + \Omega_{r_{0}}H_{0}^{2}a^{-4} + \left(\alpha H^{2}+\beta \dot{H}\right)^{1-\frac{1}{2}\Delta},
\end{equation}
where $\Omega_{i_{0}}H_{0}^{2}=\frac{\rho_{i_{0}}}{3M_{p}^{2}}$. In the following calculations, we shall perform a change of variable in order to align better our study with the late-time dynamics, i.e., we use the redshift $z$ instead of the cosmic time $t$ as the dynamical parameter. From its definition,
\begin{equation}\label{eq:redshift_scalefactor}
1+z=\frac{1}{a(t)},
\end{equation}
where we assumed that $a(t=0)=a_0=1$. From the above, the time derivatives can be expressed in terms of derivatives with respect to the redshift, using the following rule:
\begin{equation}\label{eq:time_transformation}
\frac{d}{dt}=-H(1+z)\frac{d}{dz}.
\end{equation}
Implementing such a change, Eq.~(\ref{eq:Model_FriedmannEq}) takes the form
\begin{equation}\label{eq:Hofz_equation}
(1+z) \frac{\beta}{2} \frac{d H^{2}}{dz} - \alpha H^{2}+ \left[H^{2}- \Omega_{m_{0}} H_{0}^{2} (1+z)^{3}- \Omega_{r_{0}} H_{0}^{2} (1+z)^{4} \right]^{\frac{2}{2-\Delta}}=0,
\end{equation}
where it is required that $\Omega_{m_{0}} + \Omega_{r_{0}} + \Omega_{\Lambda_{0}} = 1$\footnote{Specifically, we take $\Omega_{m_{0}}=0.315$ , $\Omega_{r_{0}}=3 \times 10^{-4}$,  $\Omega_{\Lambda_{0}}=0.6847$, and also $H_{0}=67.37$ km/s/Mpc \cite{Planck:2018vyg}.}, with the `0' index indicating the corresponding present ($z = 0$) value for each quantity i.e. matter density $\Omega_{m_{0}}$ and the Hubble parameter $H_{0}$.

\begin{figure}[ht]%
\centering
\includegraphics[width=0.49\textwidth]{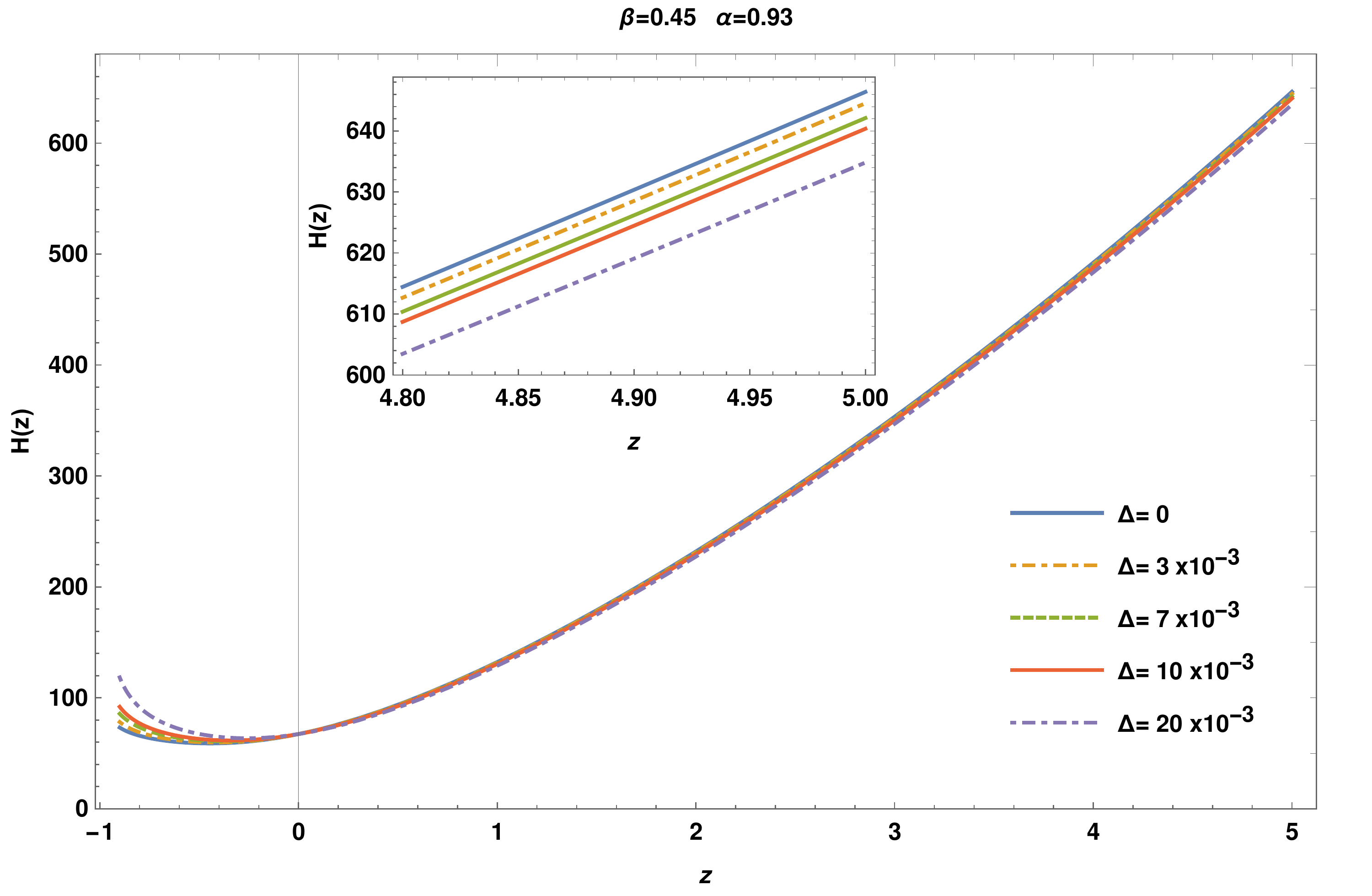}
\includegraphics[width=0.49\textwidth]{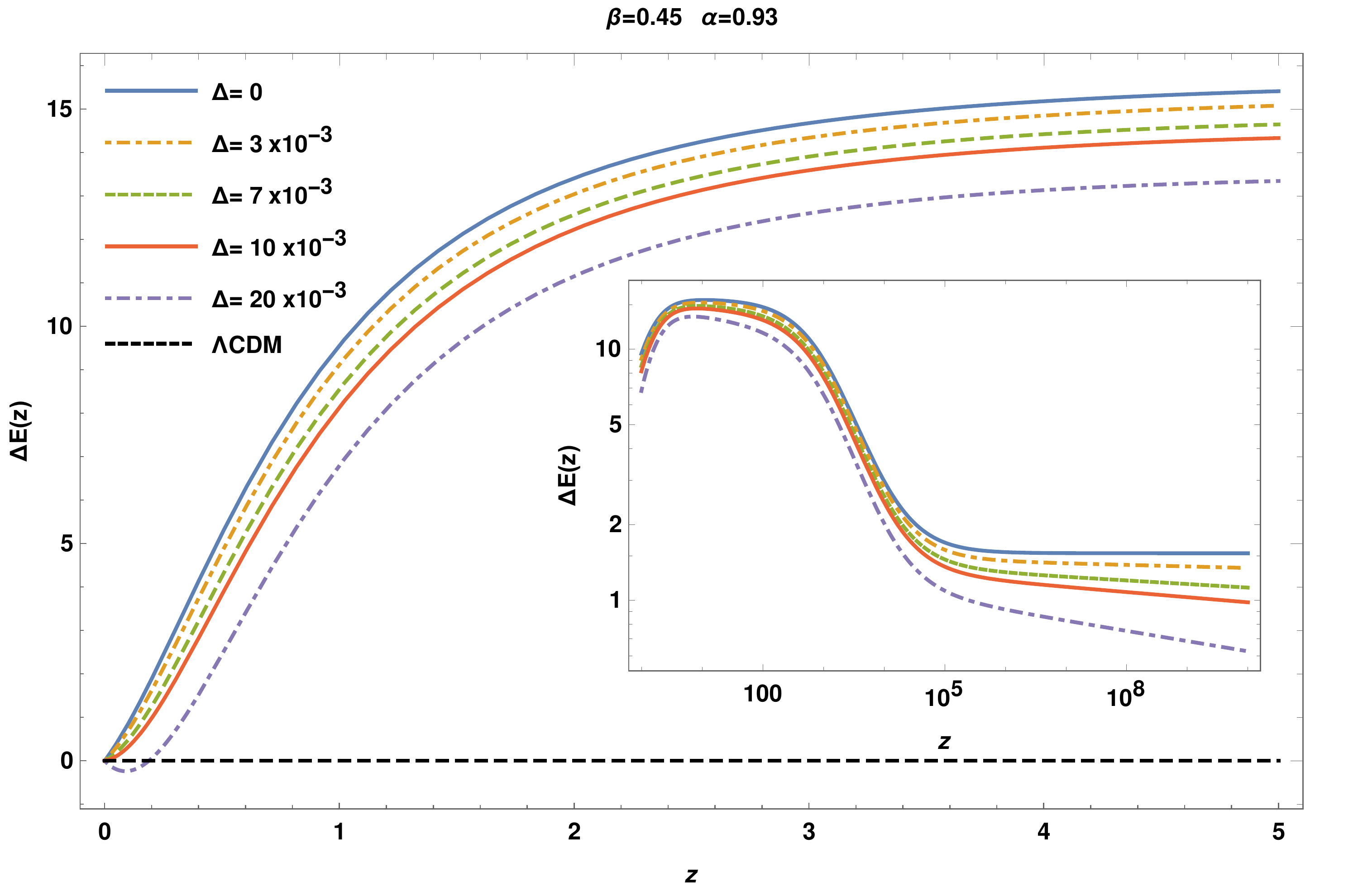}
\caption{(Left) Evolution of the Hubble parameter, $H(z)$, as a function of the redshift $z$, for different values of $\Delta$. (Right) Comparison of the proposed model against $\Lambda$CDM, for different values of $\Delta$, using the dimensionless Hubble parameter through Eq.~(\ref{eq:DeltaE}).}\label{fig:Hvsz}
\end{figure}

The complexity of Eq.~(\ref{eq:Hofz_equation}) prevents us to find an analytic solution, but it is still possible to solve it numerically and study how the Hubble parameter varies with $z$. This evolution is shown in Fig.~\ref{fig:Hvsz}, where $\alpha$ and $\beta$ are selected close to the ones reported in Ref.~\cite{Wang:2010kwa}, which considered a model with $\Omega_{r_0}=0$ and without the effect of the deformation parameter, $\Delta$. The left panel of Fig.~\ref{fig:Hvsz} shows that the evolution of $H(z)$ is barely modified when different values of $\Delta$ are considered, specially for $z \geq 0$, though increasing $\Delta$ makes $H(z)$ steadily smaller, as can be seen in the zoomed region included in this figure. Notwithstanding, notice that the impact of the deformation parameter becomes more clear in the future ($z < 0$), and the opposite effect is observed, i.e., for larger values of $\Delta$, the Hubble parameter increases. This behavior is compatible with the one obtained by other authors; for instance, see \cite{Huang:2021zgj,Rani:2021hvh}. 

It is possible to compare the model proposed here against the $\Lambda$CDM standard model using the dimensionless Hubble parameter $E(z) = H(z) / H_0$, and calculating
\begin{equation}\label{eq:DeltaE}
    \Delta E(z) = 100 \times \left[ \frac{E(z)}{E(z)_{\Lambda\rm{CDM}}} -1 \right],
\end{equation}
\\
which, by definition, is zero for the $\Lambda$CDM model (black-dashed line in the right plot of Fig.~\ref{fig:Hvsz}).

\begin{figure}[ht]%
\centering
\includegraphics[width=0.49\textwidth]{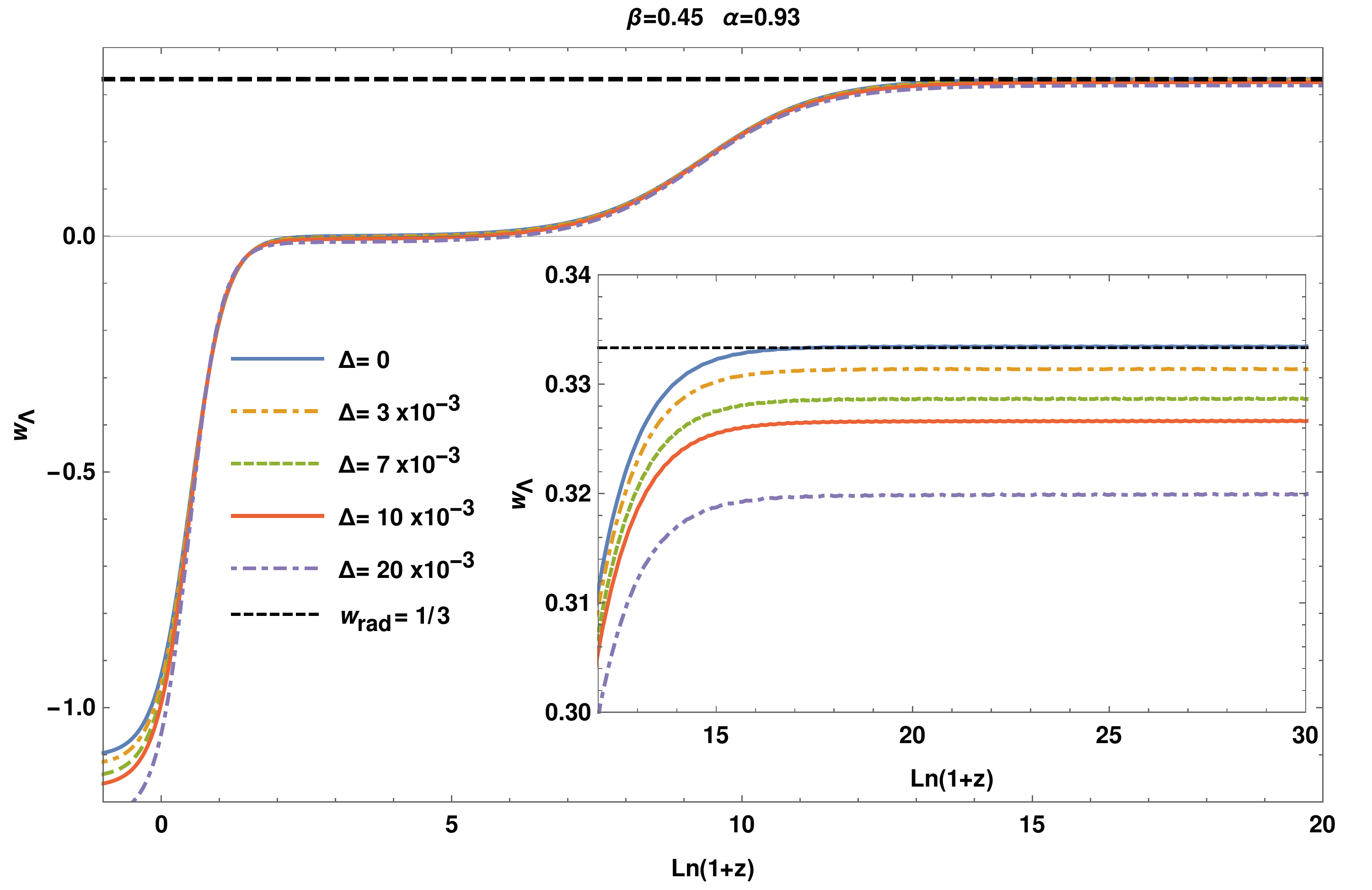}
\includegraphics[width=0.49\textwidth]{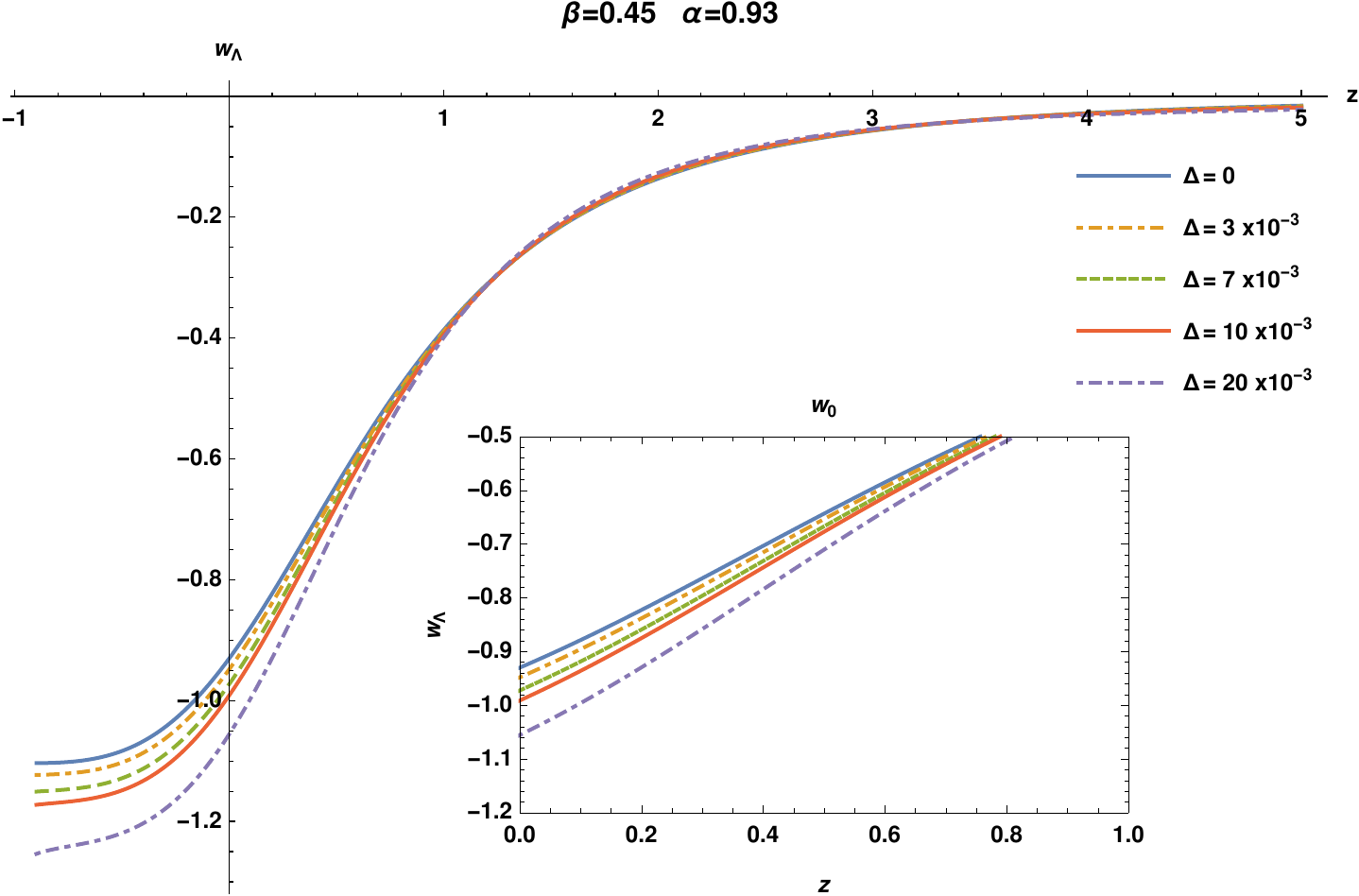}
\caption{Evolution of the EoS, $w_{\Lambda}(z)$, as a function of the redshift $z$. See text for a detailed description.}\label{fig:wDEvsz}
\end{figure}

The evolution of $\Delta E(z)$ is shown in the right plot of Fig.~\ref{fig:Hvsz}. One can see that our model gives $\Delta E(z) > 0$, indicating that $H(z)$ is larger than the $\Lambda$CDM prediction. However, increasing the deformation parameter leads negative values of $\Delta E(z)$ near the present epoch, $z=0$. In particular, for $\Delta = 2\times 10^{-4}$, there is a difference of the order of $\sim-0.5\%$ until $z \sim 0.2$, an epoch during which we observe a phantom regime with $w_{\Lambda} < -1$ (see Fig.~\ref{fig:wDEvsz} and the discussion about it, later in this section). As depicted in the inner plot, the biggest deviation from $\Lambda$CDM occurs up to early times around $z \sim 10^2-10^3$. For even earlier times ($z > 10^4$), the difference reduces reaching a plateau for some $\Delta$'s at around a few percents, and going down to $<1\%$ for large values of the deformation parameter.

\begin{figure}[ht]%
\centering
\includegraphics[width=0.6\textwidth]{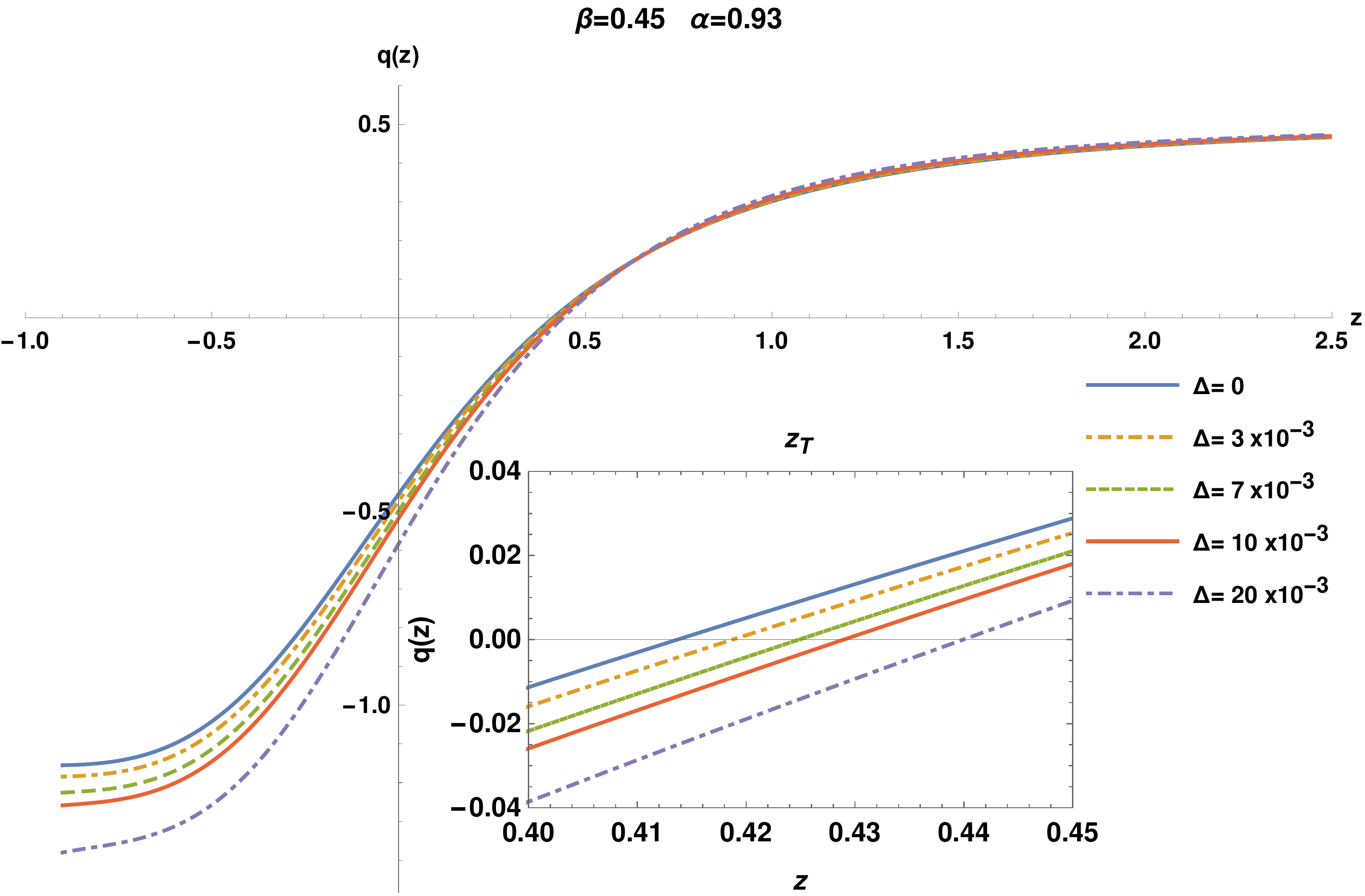}
\caption{Evolution of the deceleration parameter, $q(z)$, as a function of the redshift $z$, for different values of $\Delta$.}\label{fig:qvsz}
\end{figure}

Two other important quantities to study are the deceleration parameter, $q(z)$, and the DE equation of state (EoS), $w_{\Lambda}(z)$. The first one, the deceleration parameter, which is defined as
\begin{equation}\label{eq:qofH}
q=-1-\frac{\dot{H}}{H^2} \ \ \ \Rightarrow \ \ \ q(z)=-1+\frac{(1+z)}{2H^2}\frac{dH^2}{dz},
\end{equation}
(where we have performed the change of variable as described earlier, considering Eqs.~(\ref{eq:redshift_scalefactor}) and (\ref{eq:time_transformation})), is also computed numerically by means of solving Eq.~(\ref{eq:Hofz_equation}). The obtained evolution with the redshift is presented in Fig.~\ref{fig:qvsz}, with the same values of the parameters ($\alpha, \beta$) used before and for different values of the deformation parameter, $\Delta$.

As clearly visible in Fig.~\ref{fig:qvsz}, the deformation parameter changes markedly the evolution of the deceleration parameter. For all the cases, the transition from a decelerated  ($q > 0$) to an accelerated ($q < 0$) expansion of the universe is continuous and smooth, and the different values for the transition redshift, $z_T$ (shown in the inner panel) are compatible with those obtained from fits of recent observations \cite{Jesus:2019nnk}. We have noticed that fixing $\Delta$ and allowing the other parameters ($\alpha$ or $\beta$) to vary once at a time, produces similar results, though $z_T$ exhibits significantly larger changes in those cases, pointing to stringent restrictions to those parameters.

Regarding to the EoS parameter $w_{\Lambda}$, using the same change of variable described above, we have
\begin{equation}\label{eq:wDE}
w_{\Lambda}=-1-\frac{2}{3}\frac{\dot{H}}{H^2} \ \ \ \Rightarrow \ \ \ w_{\Lambda}(z)=-1+\frac{(1+z)}{3H^2}\frac{dH^2}{dz}.
\end{equation}
Looking at the evolution of the $w_{\Lambda}$ plotted for different values of $\Delta$ (Fig.~\ref{fig:wDEvsz}), one can see that it has an asymptotic behavior towards a radiation-type equation of state for $\Delta = 0$ at early times (left panel), though increasing $\Delta$ reduces the value of the asymptote (see the zoomed region inside the left plot). Also, at present time ($z = 0$), as $\Delta$ increases, the DE EoS exhibits (right panel of Fig.~\ref{fig:wDEvsz}) a transition from a quintessence regime ($w_{\Lambda} > -1$) to a phantom regime ($w_{\Lambda} < -1$), crossing the cosmological constant case ($w_{\Lambda} = -1$). Interestingly, for $z < 0$ (the future), there is a quintessence-phantom transformation for certain values of the parameters.

For $w_{\Lambda}$ we have also studied the effect of fixing $\Delta$ and varying $\alpha$ or $\beta$, at a time. The results indicate that the passage from a quintessence regime to a phantom one occurs when $\alpha$ ($\beta$) decreases (increases), a behavior that is characteristics of a kind of models known as Quintom \cite{Feng:2004ff}.

\begin{figure}[ht]%
\centering
\includegraphics[width=0.49\textwidth]{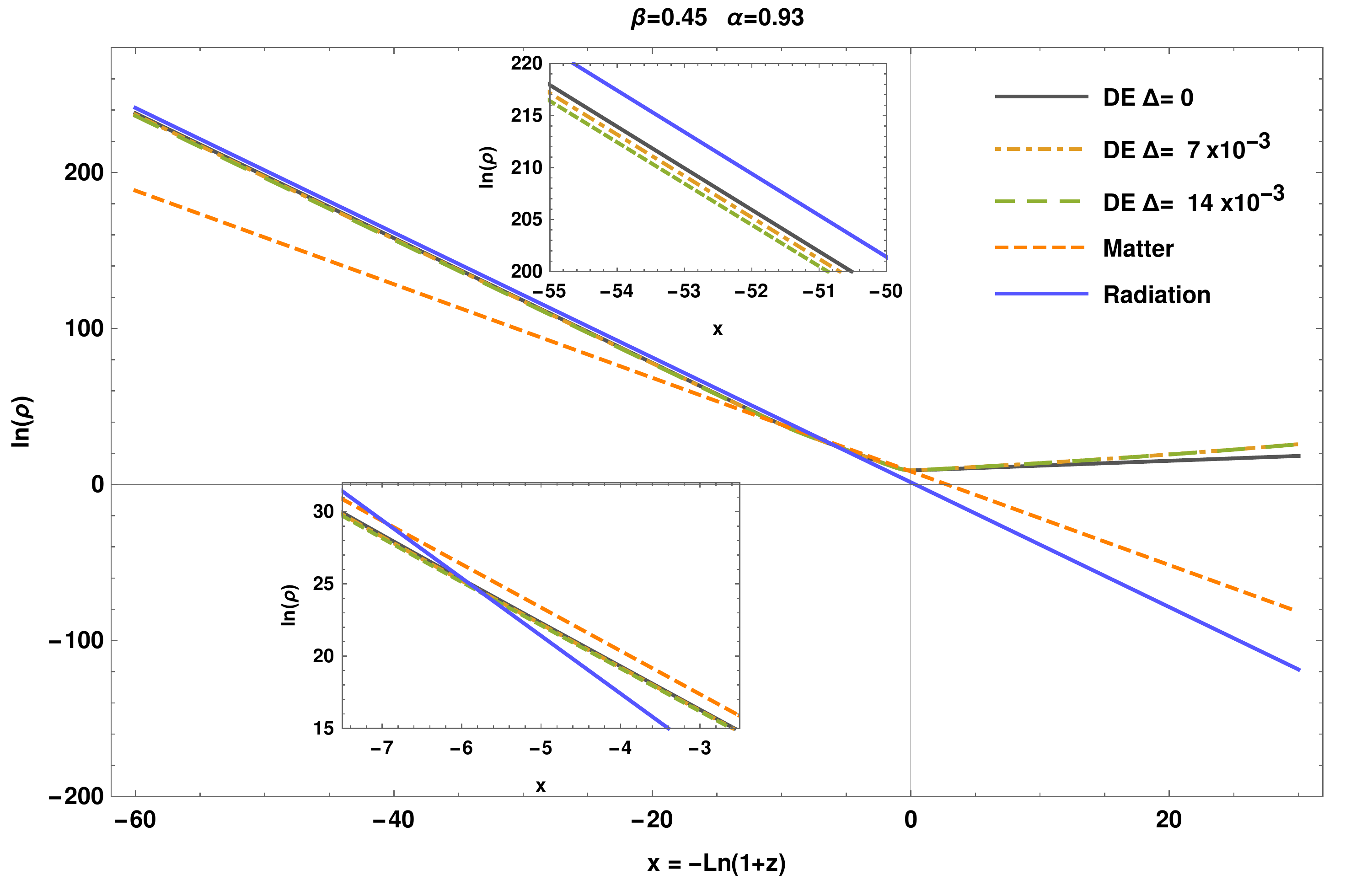}
\includegraphics[width=0.49\textwidth]{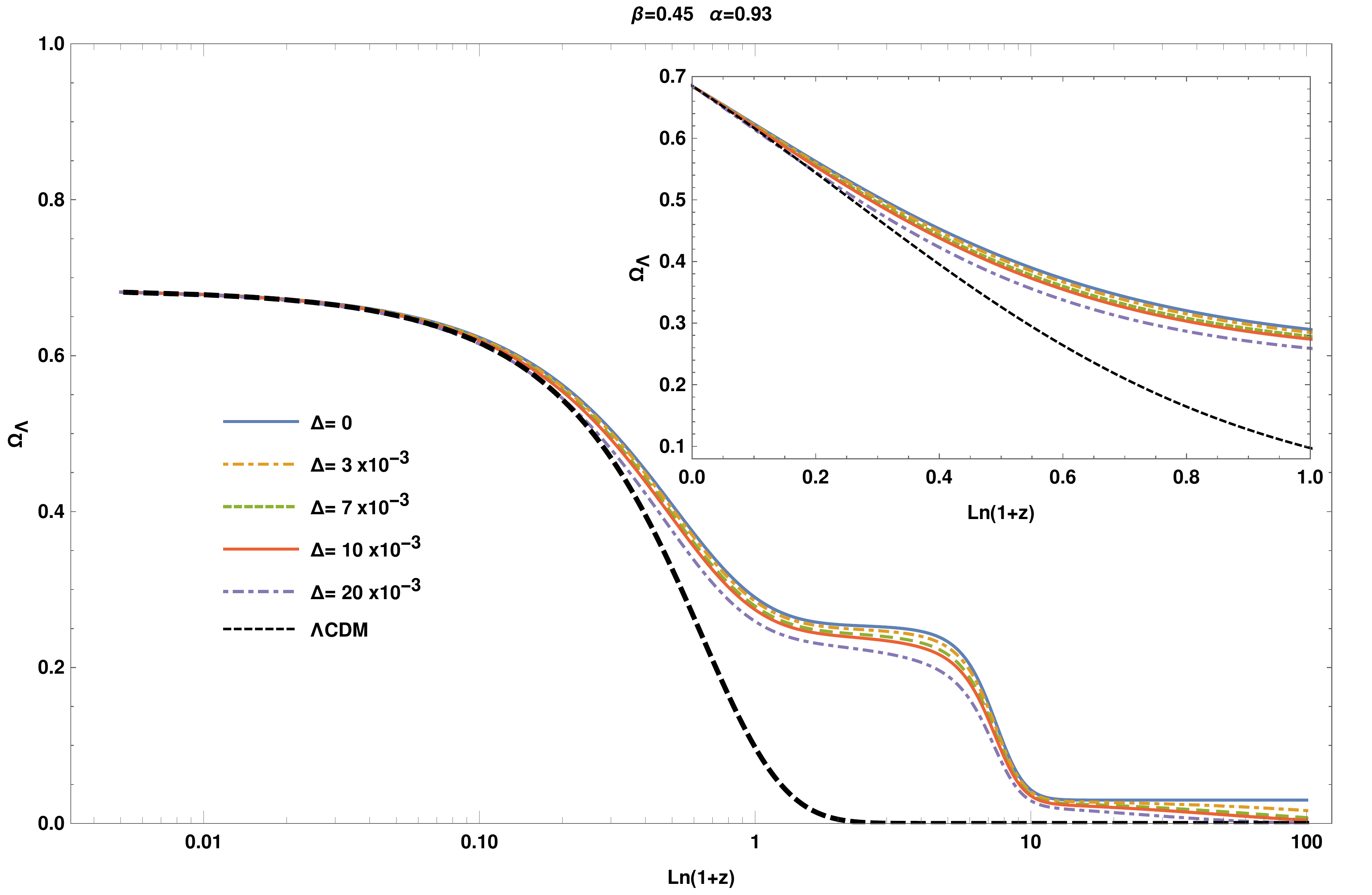}
\caption{(Left) Evolution of radiation, matter and dark energy densities as function of the number of e-folds for different values of $\Delta$. (Right) Evolution of the DE density $\Omega_{\Lambda}$ for different values of $\Delta$ and $\Lambda$-CDM, as a function of the redshift $z$.}
\label{fig:OmegaDE}
\end{figure}

In the left plot of Figure \ref{fig:OmegaDE}, the evolution of the densities involved in the model is shown for different values of $\Delta$, as a function of the number of e-folds, $x=ln(a)$. Our model exhibits an initial era of radiation dominance, followed by the non-relativistic matter and the current era of DE dominance. As can be seen, our model exhibits a non negligible contribution of the DE component at early times, when compared against radiation; this is an opposite behavior of that predicted by 
the $\Lambda$-CDM model as shown in the left panel of Figure \ref{fig:OmegaDE}, where the dark energy component, $\Omega_{\Lambda}$, is nonzero for different values of $\Delta$ in the distant past, behavior associated with the extra terms of (effective) matter and (effective) radiation added through the G-O IR cutoff to $H(z)$. According to the Holographic principle the energy density is proportional to $L^{\Delta-2}$. Since the length scale is expected to be very small during early times, the energy density generated by our model should be large enough to support the inflationary scenario. In fact, it is proposed as a possible candidate for inflation \cite{Maity:2022gdy}. 

\subsection{Stability of the model}\label{sec_stability}
\begin{figure}[ht]%
\centering
\includegraphics[width=0.6\textwidth]{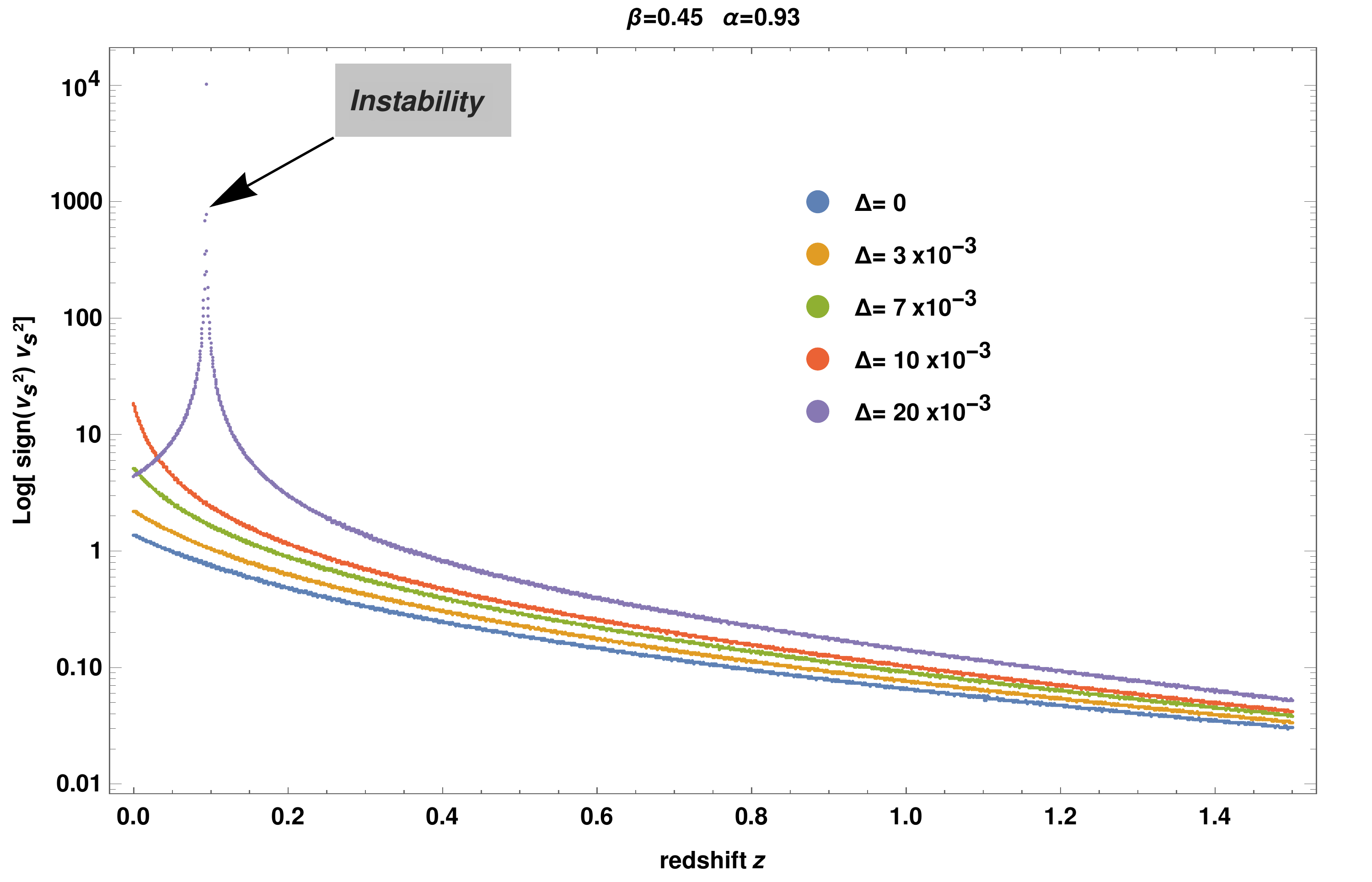}
\caption{Stability of the model under consideration trough the evolution of $v_s^2(z)$ as a function of the redshift, for different values of the deformation parameter, $\Delta$. The sharp peak in a curve indicates the presence of instabilities as $v_s^2$ becomes negative.}\label{fig:vs2vsz}
\end{figure}

\begin{figure}[ht]%
\centering
\includegraphics[width=0.49\textwidth]{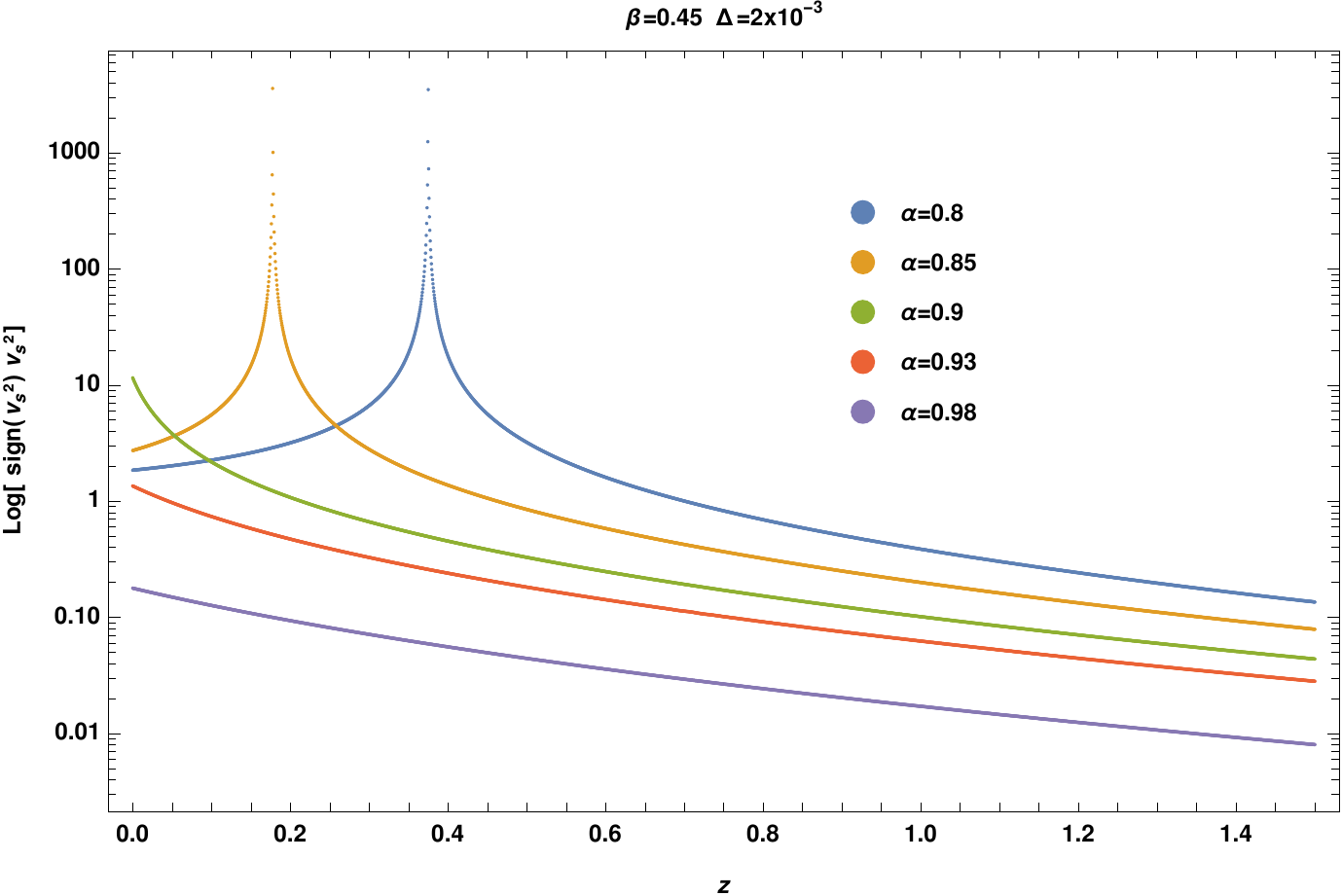}
\includegraphics[width=0.49\textwidth]{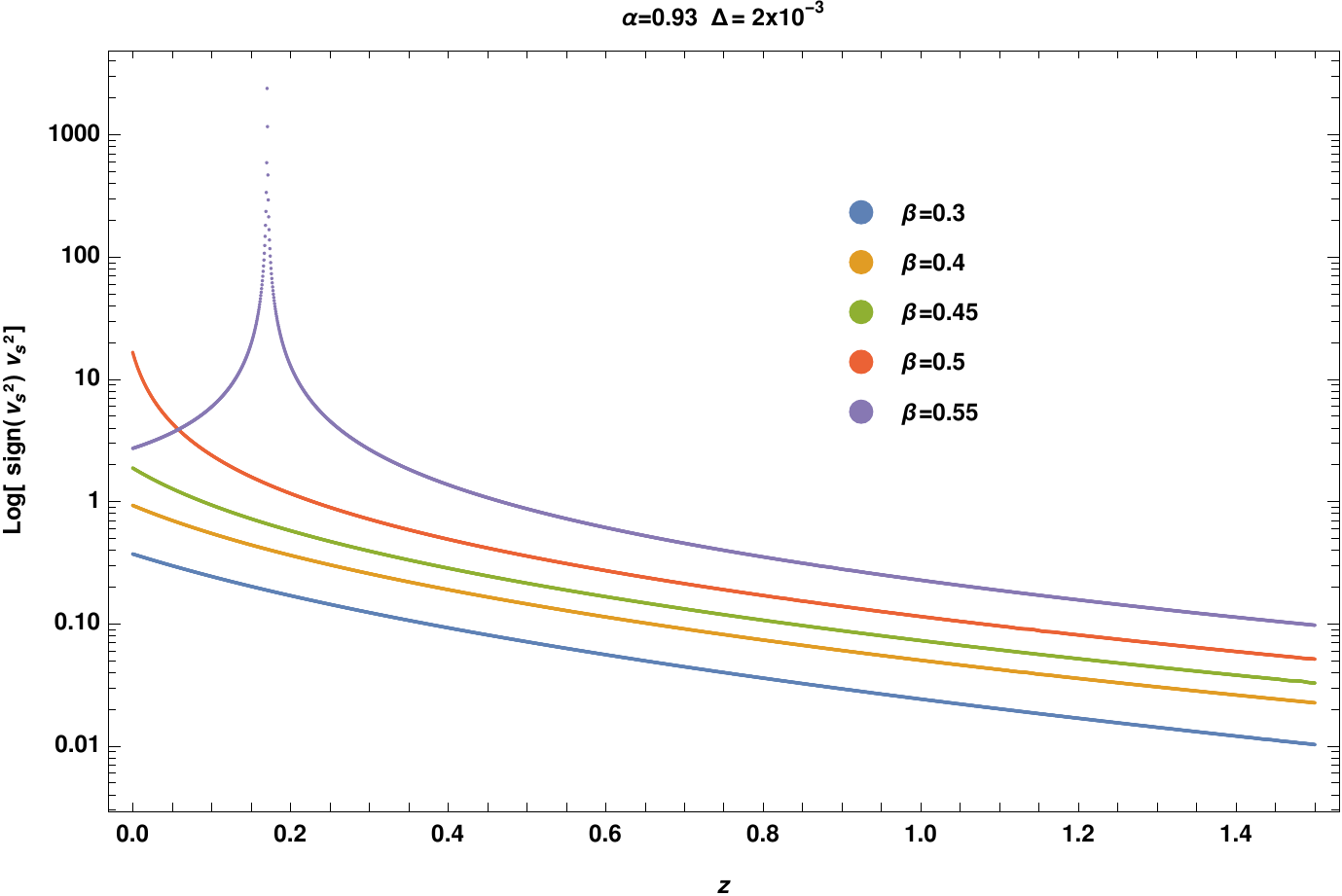}
\caption{As in Fig.~\ref{fig:vs2vsz}, but changing $\alpha$ (left) and $\beta$ (right), while the other parameters are fixed as indicated on the top of each plot.}\label{fig:vs2vsz_AB}
\end{figure}

Studying the stability of this kind of models is of paramount relevance. To do so, one can examine the square of the speed of sound, $v_s^2$, as a function of the redshift, for some values of the parameters of the model. Defined as
\begin{equation}\label{eq:vofsound}
    v_s^2 = \frac{d\,p}{d\,\rho},
\end{equation}
where $p$ is the pressure and $\rho$ the density of the fluid under investigation, a model would reveal an unstable behavior if $v_s^2 < 0$. For the model considered in this work, the evolution of $v_s^2$ is shown in Fig.~\ref{fig:vs2vsz} for different values of $\Delta$ (fixing the other parameters), and in  Fig.~\ref{fig:vs2vsz_AB} when $\alpha$ (left) or $\beta$ (right) change (fixing $\Delta$).

\begin{figure}[ht]
\begin{center}
{\includegraphics[height=0.24\textheight]{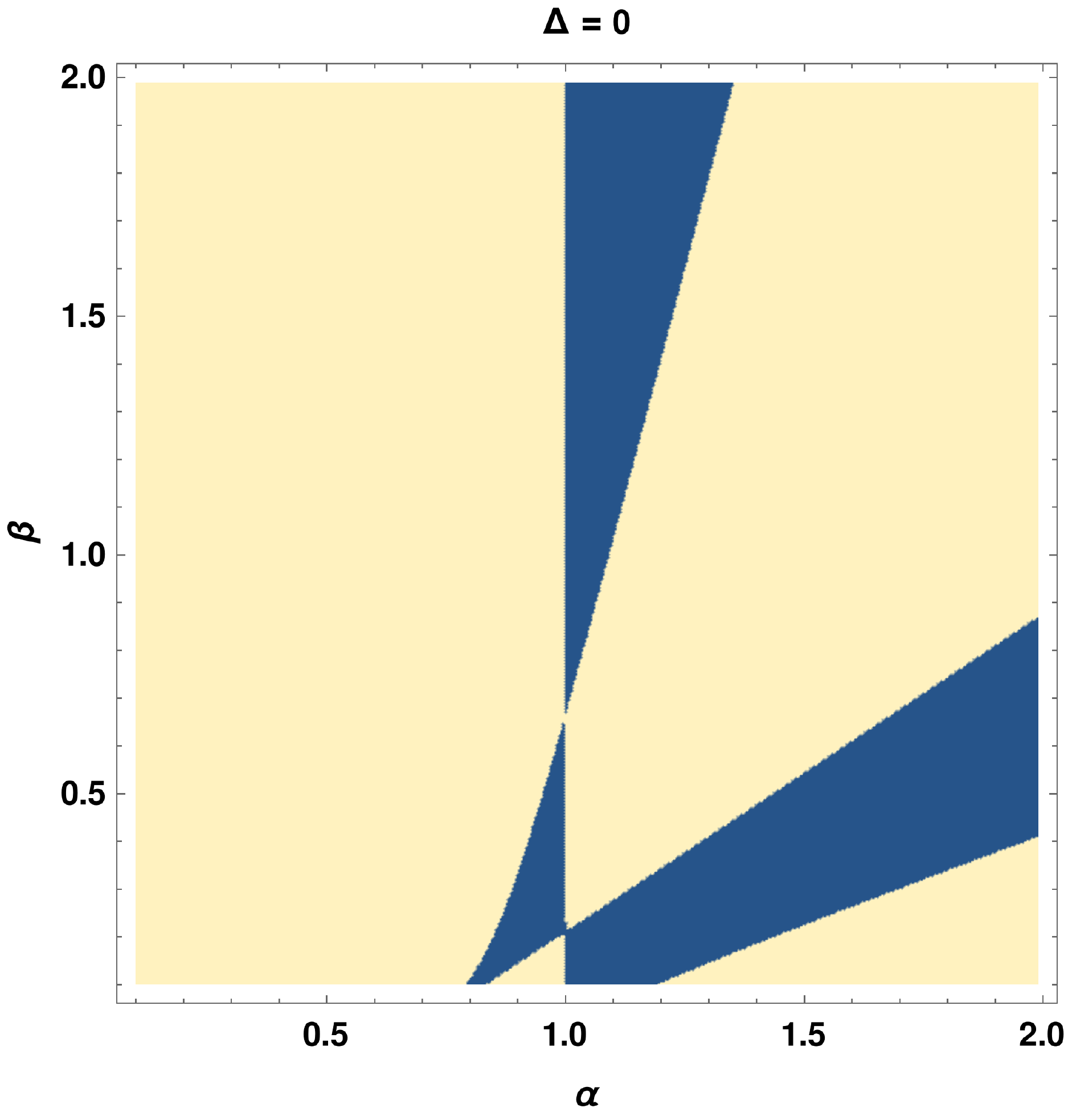}}
{\includegraphics[height=0.24\textheight]{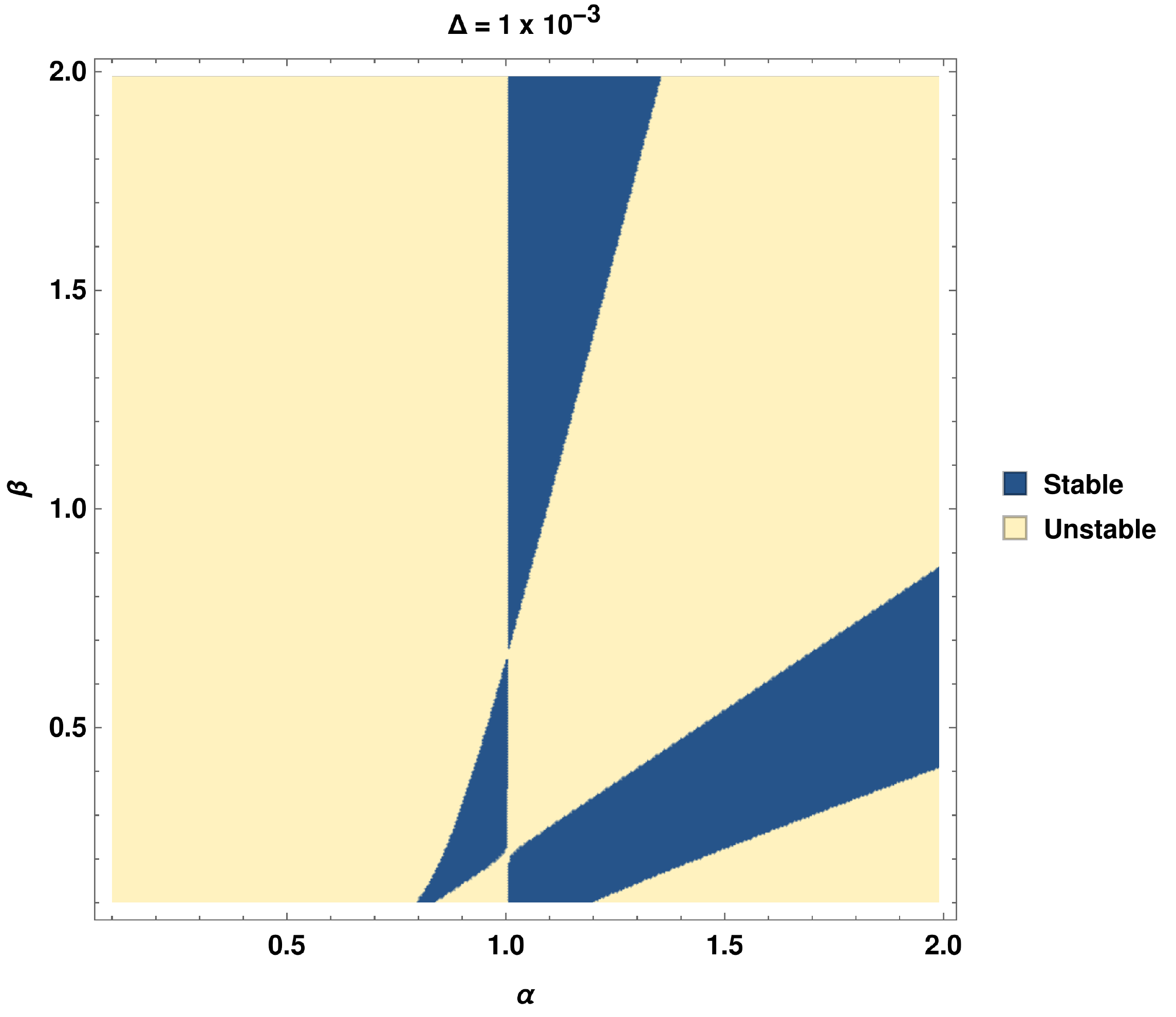}}\\
{\includegraphics[height=0.24\textheight]{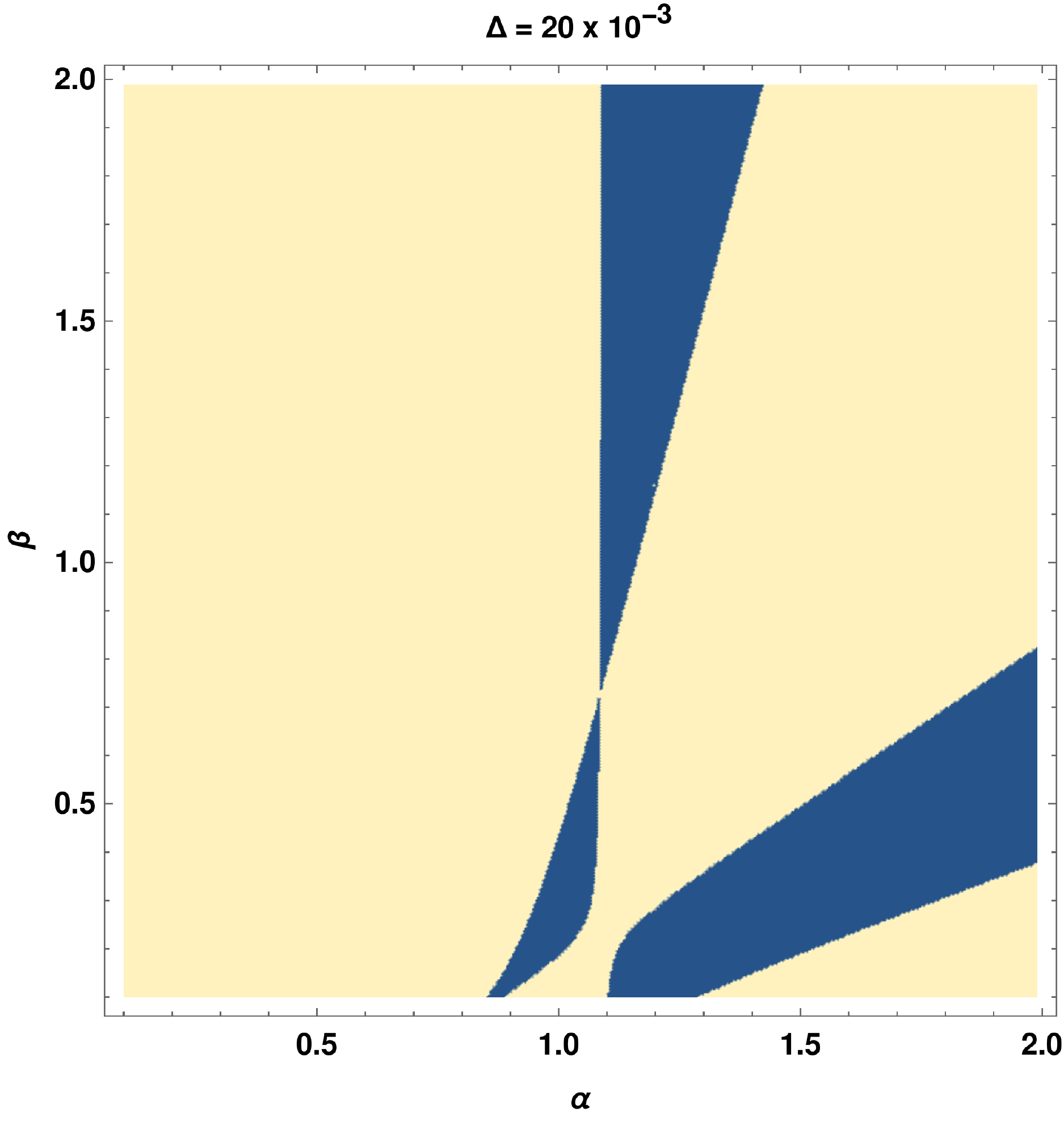}}
{\includegraphics[height=0.24\textheight]{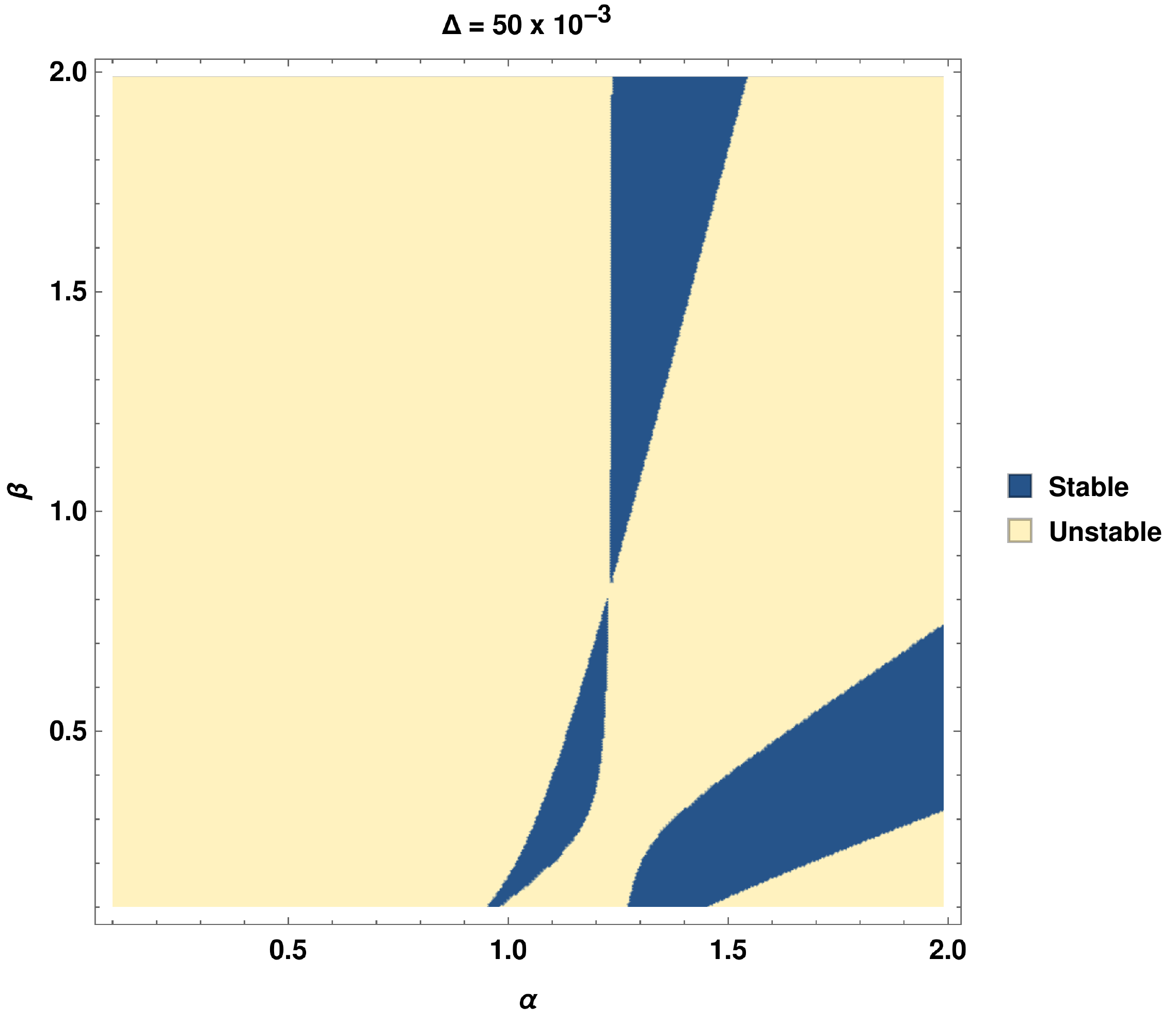}}\\
\caption{Parameter space $(\alpha,\beta)$ regions for a stable model (blue) for $\Delta = 0$ (top left), $\Delta = 1 \times 10^{-3}$ (top right), $\Delta = 20 \times 10^{-3}$  (bottom left), $\Delta = 50 \times 10^{-3}$  (bottom right). The stability criterion is $0 \leq v_{s}^{2} \leq 1$.}
\label{fig:stability}
\end{center}
\end{figure}

Because of the logarithmic scale of the vertical axis of Figs.~\ref{fig:vs2vsz} and \ref{fig:vs2vsz_AB}, the curves with a sharp peak correspond to those cases for which $v_s^2(z)$ becomes negative. That is the case for the model when $\alpha$ and $\beta$ are fixed and $\Delta$ increases, and also when $\alpha$ ($\beta$) decreases (increases), while fixing the other parameters. Notice that, nevertheless, as can be seen in Figure \ref{fig:stability}, most of the considered values for the three parameters are well within the stable region (dark blue). It is also worth mentioning that the stability zones display a reduction and displacement when $\Delta$ increase.

\section{Observational constraints}\label{sec_Fitting}
\begin{figure}[ht]
\begin{center}
{\includegraphics[height=0.28\textheight]{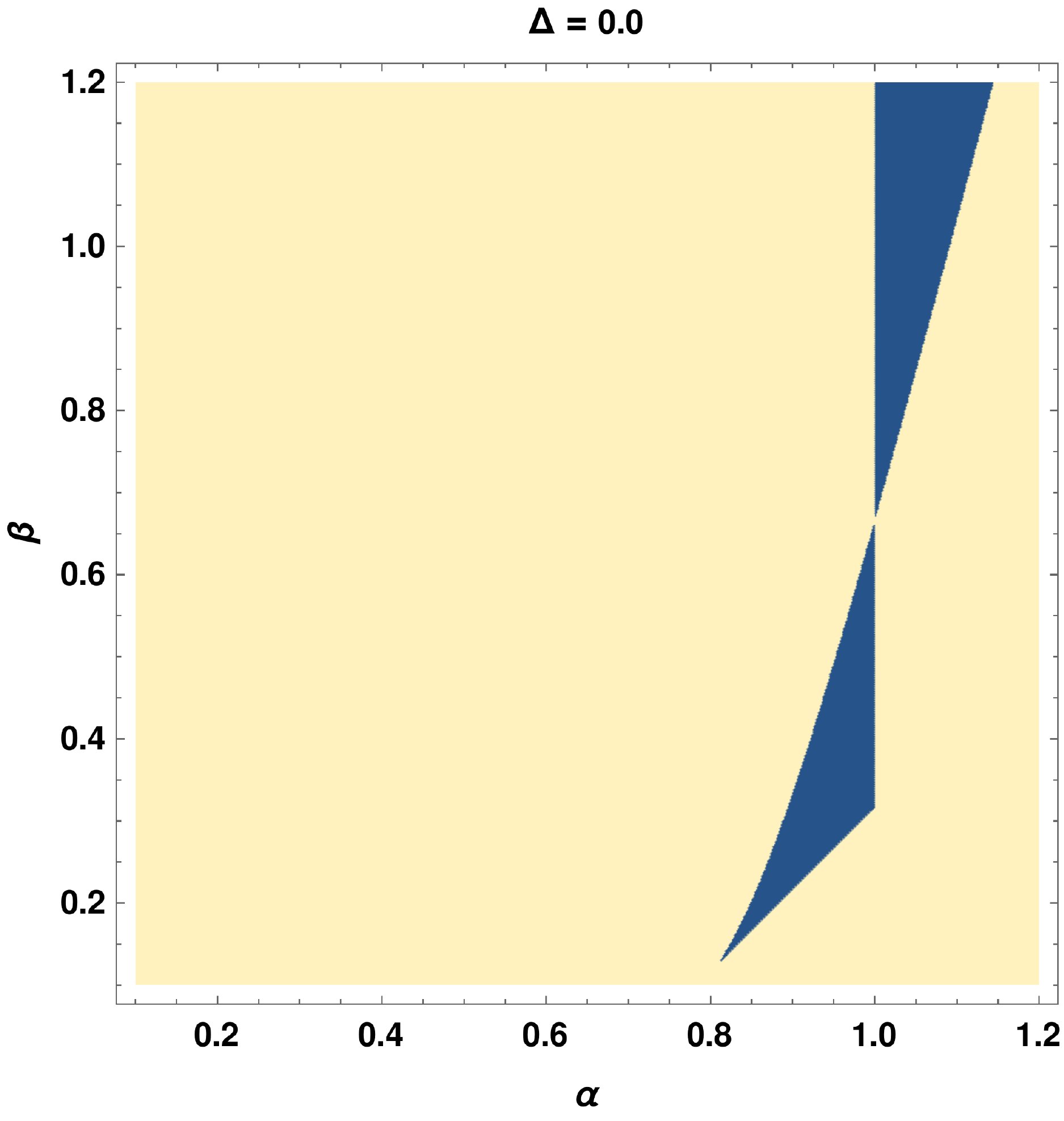}}
{\includegraphics[height=0.28\textheight]{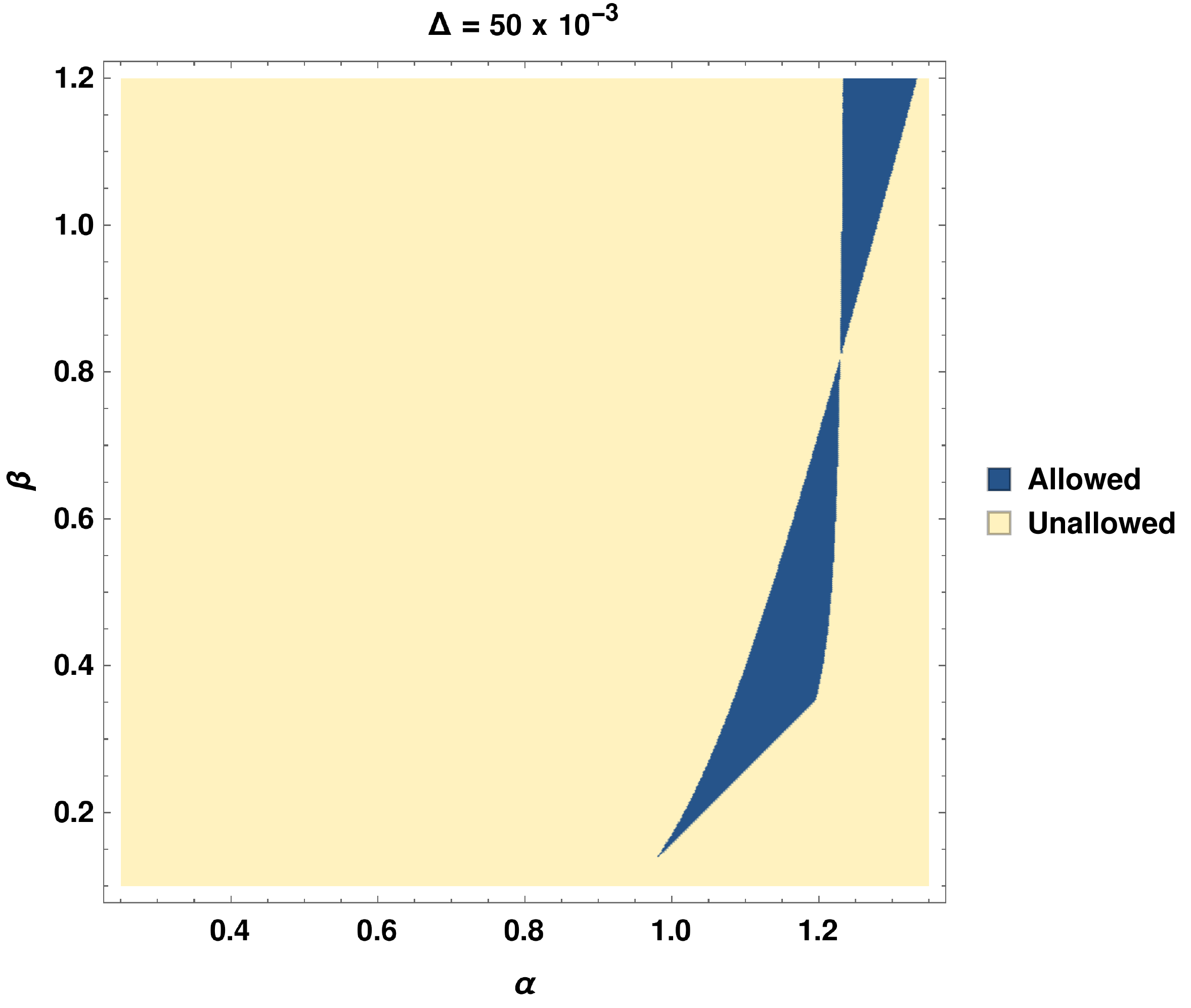}}
\caption{Parameter space $(\alpha,\beta)$ regions for a stable model (blue) for $\Delta = 0$ (left) and $\Delta = 50 \times 10^{-3}$  (right). The stability criterion is $0 \leq v_{s}^{2} \leq 1$ and we also require that $w_0 < -1/3$ and $q_0 < 0$.}
\label{fig:stabilityB}
\end{center}
\end{figure}

The thorough study described in Sec.~\ref{sec_stability} allows us to find that the stability of the model is preserved when the parameters take the values in the ranges presented in Table \ref{Tab:ranges}. This parameter space (see Fig.~\ref{fig:stabilityB}) is obtained by further requiring that $w_0 < -1/3$ and $q_0 < 0$, so that the physical results are consistent with current observations. Negative values for $\alpha$ and $\beta$ were also excluded for not complying with the imposed restrictions. 
\begin{table}[ht]
\begin{center}
\begin{minipage}{0.35\textwidth}
\caption{Parameter ranges considered for the fit.}
\label{Tab:ranges}
\begin{tabular}{@{}cc@{}}
\toprule
Parameter & Considered range  \\
\midrule
$\alpha$ & [0.7 , 1.0] \\ 
$\beta$  & [0.3 , 0.9] \\
$\Delta$ & [0.0 , 30 $\times 10^{-3}$ ]\\
\botrule
\end{tabular}
\end{minipage}
\end{center}
\end{table}

\begin{figure}[ht]%
\centering
\includegraphics[width=0.675\textwidth]{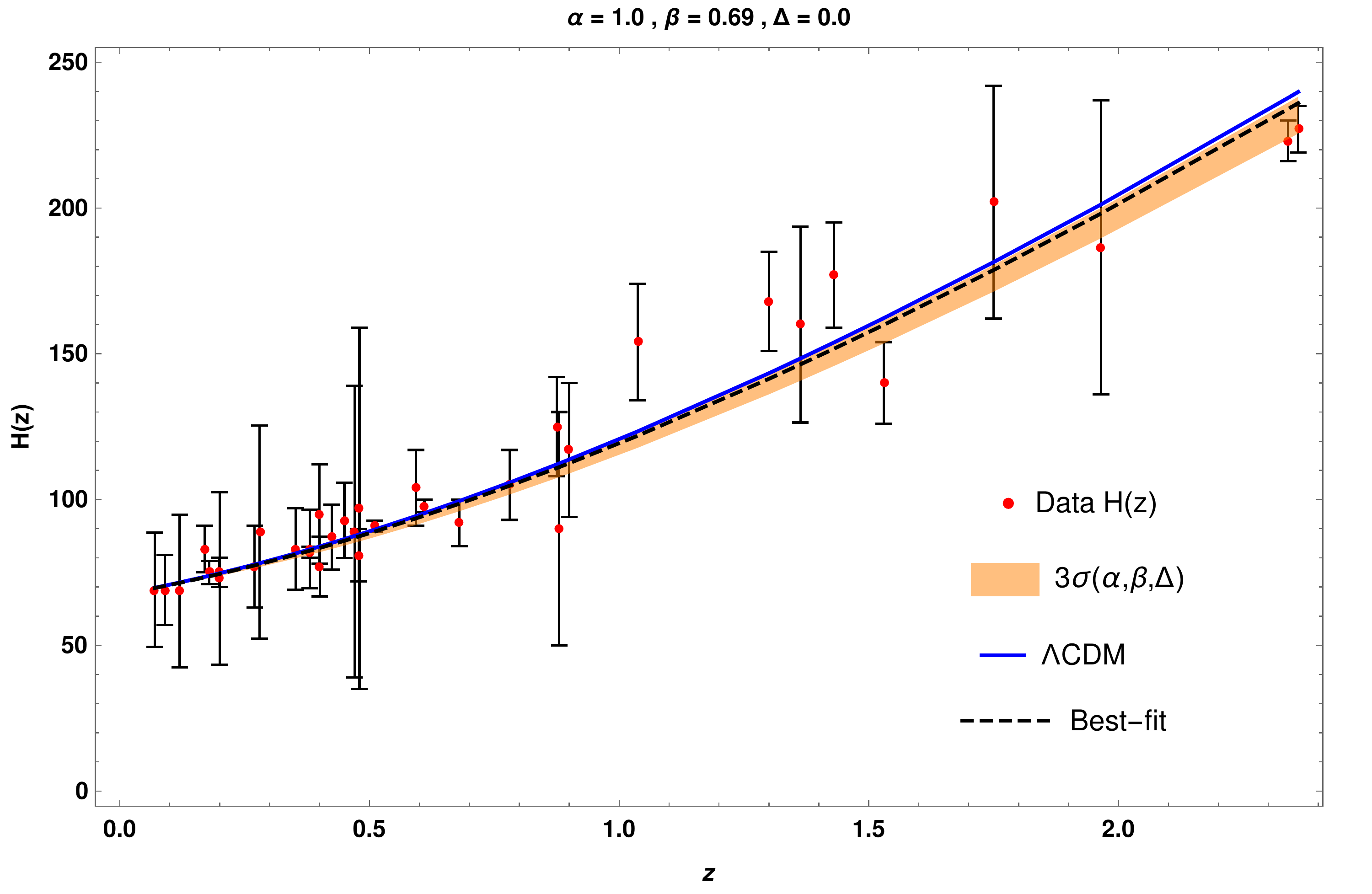}
\caption{Comparison of $H(z)$ predicted by our model using the best fit values of the parameters shown in Table~\ref{Tab:BestFit} (black-dashed line), against the observational data from \cite{Cao:2021uda}. The shadowed-orange area gives the prediction obtained when the parameters are varied within their $3\sigma$ allowed region, and the blue-full line is the $\Lambda$CDM prediction.}\label{fig:FittedData_3Sig}
\end{figure}

The analysis reported in this section is performed by comparing the values of $H(z)$ as calculated with our model against recent observations, in order to fit the parameters of the model. The specific observational data consists of 36 measurements as reported in Ref.~\cite{Cao:2021uda} and shown here in Fig.~\ref{fig:FittedData_3Sig}. The statistical analysis is implemented through a standard $\chi^2$ function defined as
\begin{equation}\label{eq:chi2}
\chi^{2}_{H(z)}(\theta) = \sum_{i=1}^{36} \frac{\left( H_{th}(\theta,z_{i}) - H_{obs}(z_{i})\right)^{2} }{\sigma^{2}_{i}},
\end{equation}
where $\theta = (\alpha,\beta,\Delta)$ is the vector of parameters to be fitted, $H_{th}$ and $H_{obs}$ are the predicted and measured values of the Hubble parameter, for different $z_i$, respectively, and $\sigma_i$ is the error associated to observation $i$.

\begin{figure}[htp]
    \centering
    \includegraphics[width=0.7\textwidth]{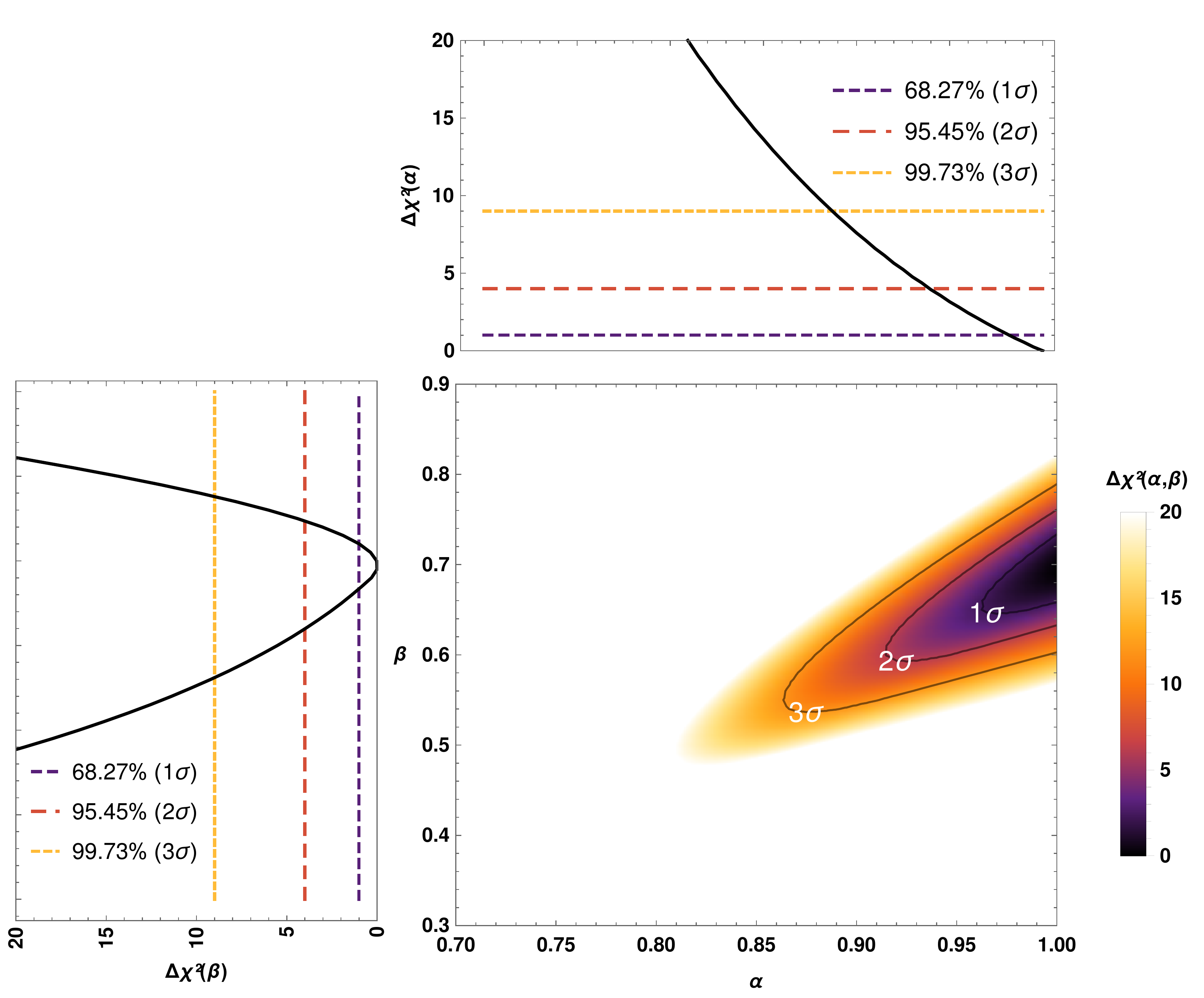}
    \caption{Allowed regions for the parameter space ($\alpha,\beta$) at 68,27\%, 95,45\%, 99,73\% C.L., as obtained from the fit.}
    \label{Fig:marginAB}
\end{figure}

\begin{figure}[htp]
    \centering
    \includegraphics[width=0.7\textwidth]{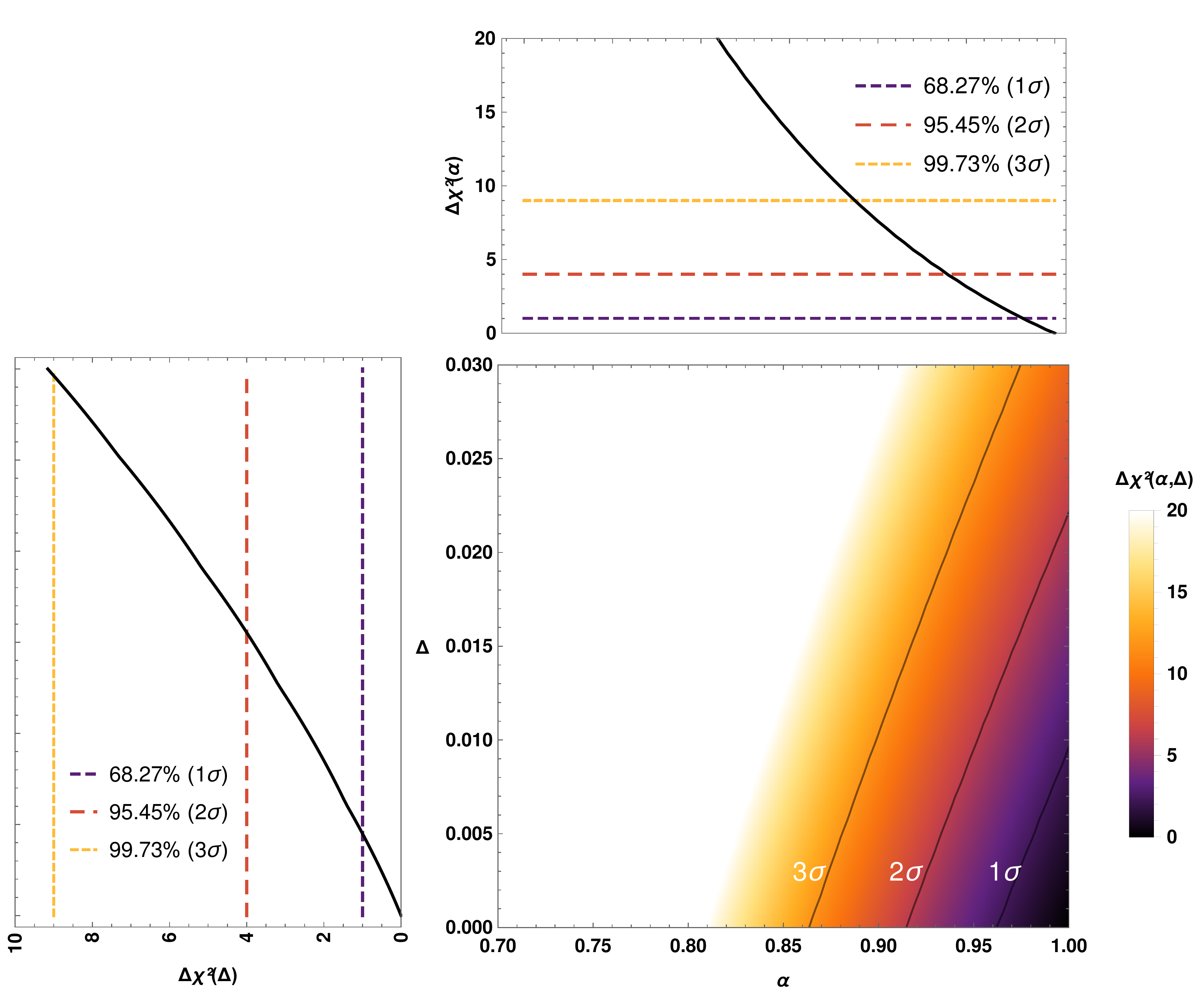}
    \caption{Allowed regions for the parameter space ($\alpha,\Delta$) at 68,27\%, 95,45\%, 99,73\% C.L., as obtained from the fit.}
    \label{Fig:marginAD}
\end{figure}

\begin{figure}[htp]
    \centering
    \includegraphics[width=0.7\textwidth]{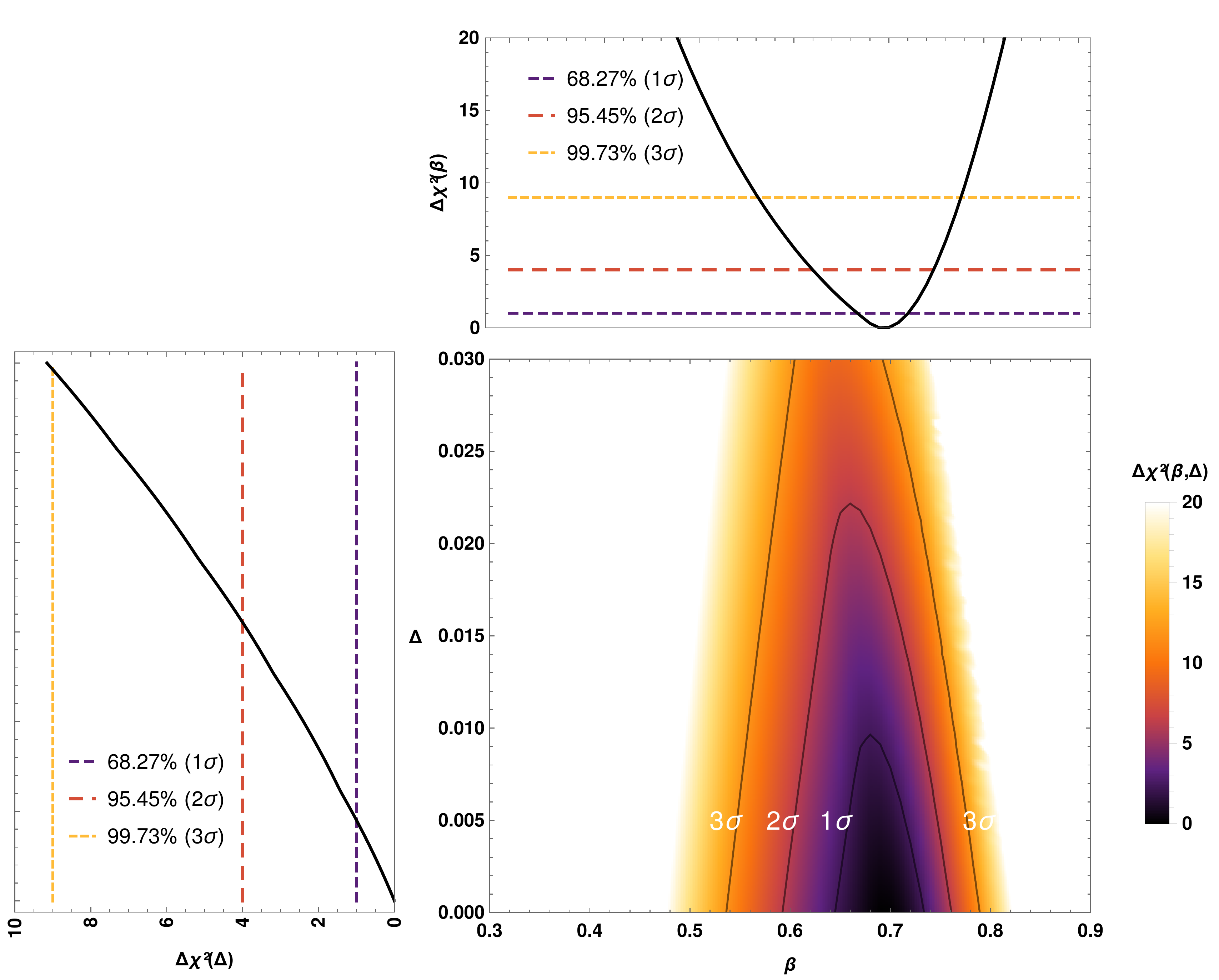}
    \caption{Allowed regions for the parameter space ($\beta,\Delta$) at 68,27\%, 95,45\%, 99,73\% C.L., as obtained from the fit.}
    \label{Fig:marginBD}
\end{figure}

In Figures \ref{Fig:marginAB}, \ref{Fig:marginAD} and \ref{Fig:marginBD} we show the results of the analysis. In each Figure, the central 2D contours correspond to the allowed regions at $1\sigma$, $2\sigma$, $3\sigma$\footnote{Corresponding to 68.27\%, 95.45\% and 99.73\% confidence level (C.L.), respectively \cite{ParticleDataGroup:2020ssz}.} of the respective (2D) parameter space (i.e., a pair of parameters of the model), resulting from the marginalization of the third parameter (not shown in each case). The 1D allowed intervals for each parameter are also shown in the top and left plots.

The three set of Figures are consistent among each other in the sense that the obtained allowed regions for each parameter are clearly compatible, and the best fit (i.e., the set of parameters that minimize $\chi^2_{H(z)}$, Eq.~(\ref{eq:chi2})) is found at $(\alpha,\, \beta,\, \Delta) = (1.0,\,0.69,\,0.0)$, with the overall allowed intervals as shown Table \ref{Tab:BestFit}. The contour plots suggest that the combined observational data of $H(z)$ provide a strong constraint on $\beta$, while the most likely value of $\Delta$ is the minimum ($\Delta=0$) within the region of $1\sigma$, in agreement to what was recently reported in Ref.~\cite{Anagnostopoulos:2020ctz}, for which the entropy-area relation is consistent with the Bekenstein-Hawking entropy.
\begin{table}[ht]
\begin{center}
\begin{minipage}{0.75\textwidth}
\caption{Best fit and allowed regions for the parameters of the model.}
\label{Tab:BestFit}
\begin{tabular}{@{}ccccc@{}}
\toprule
Parameter & Best fit & $1\sigma$ & $2\sigma$ & $3\sigma$  \\
\midrule
$\alpha$ & 1.00 & (0.98, 1.02)  & (0.94, 1.06)  & (0.89, 1.11)  \\ 
$\beta$  & 0.69 & (0.67, 0.72)  & (0.62, 0.75)  & (0.56, 0.78)  \\
$\Delta$ & 0.00 & (0.00, 0.004) & (0.00, 0.016) & (0.00, 0.029) \\
\botrule
\end{tabular}
\end{minipage}
\end{center}
\end{table}

For comparison, the Hubble parameter as a function of the redshift predicted by our model is represented by the black-dashed line in Fig.~\ref{fig:FittedData_3Sig}, where the observational data are also drawn, together with the prediction of the $\Lambda$CDM model (blue-full line). The shadowed (orange) region represents the prediction within the $3\sigma$ variation of the parameters of our model (see Table \ref{Tab:BestFit}). Our proposal not only presents a similar behavior compared to $\Lambda$CDM, but more importantly, it fits very well to the observational data, with $\chi^2/dof = 22.6/33$\footnote{$dof$ stands for \emph{degrees of freedom}, calculated as the difference between the number of data points and the number of free parameters.}.

In addition, implementing our model with the best fit values for the parameters (Table \ref{Tab:BestFit}), we can compute the current value of the EoS parameter, obtaining $w_{0}= -1.016^{+0.009}_{-0.005}$, which is consistent with the cosmological constant ($w_{0}= -1$) at $1.8\sigma$. An estimate of the transition redshift is also computed, and we get $z_{t}=0.658^{+0.020}_{-0.012}$, as well as for the deceleration parameter at present times, $q_{0}=-0.543^{+0.009}_{-0.005}$. These results are comparable to previously reported ones, $z_{t}=0.65^{+0.10}_{-0.07}$ \cite{Mamon:2017rri} and  $q_{0}=-0.6401\pm0.187$ \cite{Capozziello:2018jya}, respectively. Furthermore, we estimate the age of the Universe to be $t_{0}=14.24^{+0.11}_{-0.06}$ Gyr, relatively close to the value reported by Planck 2018 \cite{Planck:2018vyg}, $t_{0}=13.79\pm0.02$ Gyr. 

\section{Conclusions}\label{sec_Conclusions}
{In this work we have studied the effects on the cosmological evolution of the universe for late times, taking into account the Granda--Oliveros infrared cutoff in the recently introduced Barrow Holographic Dark Energy model. First, the evolution of $H(z)$ was analyzed (see left panel of Fig.~\ref{fig:Hvsz}) and we found that the evolution of $H(z)$ is hardly distinguishable when different values of $\Delta$ are used, specially for $z \geq 0$; however, it was noted that, increasing $\Delta$ makes $H(z)$ smaller. Also, the differences against the $\Lambda$CDM prediction (right panel of Fig.~\ref{fig:Hvsz}) are significantly reduced at earlier times. Secondly, studying the deceleration parameter (Fig.~\ref{fig:qvsz}) we showed that our model predicts an accelerated expansion regime of the universe at late times. We also verified that increasing $\Delta$ causes the EoS parameter to exhibit a transition from quintessence to phantom regimes, both at late and future times (see right panel of Fig.~\ref{fig:wDEvsz}), while at early times the asymptotic behavior towards a radiation-type EoS is reduced for larger $\Delta$ (see left panel of Fig.~\ref{fig:wDEvsz}). Moreover, our model exhibits the know eras of dominance (radiation, cold matter and DE), in which at early times the DE component may not be negligible compared to radiation (see Fig.~\ref{fig:OmegaDE}). 

We studied the zones of stability before perturbations (see Fig.~\ref{fig:vs2vsz} and \ref{fig:vs2vsz_AB}) from the early period to the present time, however, we found that these regions exhibit a reduction and displacement when $\Delta$ increases (see Fig.~\ref{fig:stability}), and excludes the negative values of $(\alpha,\beta)$ as that part of the parameter space would not be consistent with physical results (see Fig.~\ref{fig:stabilityB}). 

At last, we set constraints to the model parameters (Table \ref{Tab:BestFit}) from a fit to observational data of $H(z)$, which suggests that the growth of $\Delta$ may not favor the model, and provide a rather strict restriction on $\beta$ (see Figs.~\ref{Fig:marginAB} and \ref{Fig:marginBD}). From the fit, $\Delta = 0.0$ is found to be the most likely value (see Fig.~\ref{Fig:marginAD} and \ref{Fig:marginBD}), for which the Bekenstein-Hawking relation is favored. However, it would be necessary to use a more complete set of observational  data to obtain a best fit on the parameters of the model, e.g., using updated measurements from the dynamics of the expansion of the universe, $H(z)$, and the growth rate of cosmic structures, $[f\sigma_8](z)$, or $H(z)$ and data from Supernovae (SNIa) Pantheon sample, among others, but that kind of analysis is beyond the scope of this work and could be addressed later.}

\bmhead{Acknowledgments}
A.O. acknowledges financial support from  Patrimonio Aut\'onomo--Fondo Nacional de Financiamiento para la Ciencia, la Tecnolog\'ia y la Innovaci\'on Francisco Jos\'e de Caldas (MINCIENCIAS-COLOMBIA) Grant No.~110685269447 RC-80740-465-2020, projects 69723 and 69553. This work has been partially performed using the software Mathematica.\\

\section*{Statements and Declarations}
\begin{itemize}
\item Availability of data and materials. All data that have
been used in our analysis have already been freely released and have
been published by the corresponding research teams. In our text we
properly give all necessary References to these works, and hence no
further data deposit is needed.
\end{itemize}

\end{document}